\documentclass[12pt]{article}
\usepackage{graphicx}
\usepackage{cite}
\usepackage{amsmath}
\usepackage[margin=1in]{geometry}  

\begin{document}

\title{Democratic Thwarting of Majority Rule in opinion dynamics:
1. Unavowed Prejudices versus Contrarians}

\author{ Serge Galam\thanks{serge.galam@sciencespo.fr} \\
CEVIPOF - Centre for Political Research, Sciences Po and CNRS,\\
1, Place Saint Thomas d'Aquin, Paris 75007, France}

\date{October 8, 2024}

\maketitle

\begin{abstract}

I study the conditions under which a democratic dynamics of a public debate drives a Minority-to-Majority transition. A  landscape of the opinion dynamics is thus built using  the Galam Majority Model (GMM) in a 3-dimensional parameter space for three different sizes $r=2, 3, 4$ of local discussing groups.  The related parameters are $(p_0, k, x)$, the respective proportions of initial agents supporting opinion A, unavowed tie prejudices breaking in favor of opinion A, and contrarians. Combining $k$ and $x$ yields unexpected and counterintuitive results. In most part of the landscape the final outcome is predetermined with a single attractor dynamics independently of the initial supports for the competing opinions. Large domains of $(k, x)$ values are found to lead an initial minority to turn majority democratically without any external influence. A new alternating regime is also unveiled in narrow ranges of extreme proportions of contrarians. The findings indicate that the expected democratic character of free opinion dynamics is indeed rarely satisfied. The actual values of $(k, x)$ are found to be instrumental to predetermine the final winning opinion. Therefore, the conflicting challenge for the predetermined opinion to loose,  is to modify these values appropriately to become the winner.  However, developing a model which could help manipulating public opinion rises ethical questions. The issue is discussed in the conclusion.

\end{abstract}


\newpage
\section{Introduction}

In most today democratic countries the demand for more direct democracy is growing substantially. In particular, launching public debates and referendums to address  major contemporary societal issues is seen as paramount to ensure subsequent democratic choices, as opposed to decisions taken by political decision-makers, who are often perceived as being disconnected from the reality experienced by citizens citizens \cite{refe}.

A follow up collective majority voting then achieves the democratic process of selection of appropriate policies to deal with the essential topics contemporary societies are facing. With majority rule voting between two competing choices, in principle, one ballot difference is enough to determine the winner of a binary voting.  Clearly, that never happens for large scales voting. Nevertheless, the outcome can be very tight as observed with hung elections \cite{cont}.

In this paper I question the validity of what  I quote as the ``belief" that an open public debate brings out the choice that is supported by the majority, i.e., more than half of the community concerned by the issue at stake. 

To be precise, I do not question the democratic nature of a majority rule voting, I claim that the debate taking place prior to the vote,  does modify the initial majority of individual choices along invisible and unconscious biases most often favorable to the initial minority but not always. I thus focus on unveiling the conditions under which a public debate twists ``naturally" the initial majority-minority balance by implementing a Minority-to-Majority transition. When more than two choices are competing the fairness and flaws of majority voting could be also at stake as discussed long time ago \cite{cond}. 

The present work subscribes to the emerging and active field of sociophysics  \cite{brazil, frank, book, bikas}. Sociophysics explores and  tackle social, political and psychological phenomena by adopting a physicist-like approach  \cite{inter, nun1}. The goal is not to substitute for the social sciences but to create a new hard science by itself \cite{phys1, phys2, phys3, phys4, phys5, phys7, phys8, phys10}. 

Thanks to its universal features sociophysics allows dealing with a rather large spectrum of different issues \cite{lima2, zim1, zim2, vaz, r2, r3, r4, r5, ahad2, memo, polari, plos, cui, liu, banish, depo, decis, nuno3, brics, roni3, rebel, shen, grabish, nuno4, nuno5, roni1, body}. Among them, the study of opinion dynamics has been particularly prominent \cite{mobile1, hu, tot2, sen, zan, che, dispa, roni2, andre1, ahad1, chen, x1, x2, x3, x4, x5}.  Many papers use binary variables  \cite{tot1, mala, mak, red, kas1, bol, bag, car,  kas2, mar1,mar2, igl, fas, gim, iac, bru, paw, we, pol} with a few ones getting to three or more discrete opinions  \cite{mobilla, celia, andre2, malarz, mobile2, entro, trade, tim}. Continuous variables are also considered \cite{r1, p11}.

Indeed, the Galam Majority Model (GMM) has already highlighted a phenomenon of a Minority-to-Majority transition, which goes unnoticed being unconscious and invisible to the involved agents \cite{stat, chop, min}. In particular, the GMM has unveiled the drastic biasing effect of some psychological traits including tie breaking prejudice \cite{min}, contrarian behavior \cite{cont}, and stubbornness \cite{mosc1, stub}.  These three traits were shown to produce different types of polarization, respectively unanimity, coexistence, and rigidity \cite{pola},

In addition, combining $k$ and $x$ was found to yield unexpected and counterintuitive results. The investigation was restricted to the subspace $(k, x < 0.50)$ for update groups of size 4. Yet, related findings allowed me to propose an alternative novel scheme to block the propagation of fake news without banning them  but using sequestration instead \cite{fake}. 

Here I explore the full subspace $(k, x)$ of the 3-dimensional space parameter $(p_0, k, x)$ of the opinion dynamics landscape obtained from the GMM for respective groups of size $r=2, 3, 4$. The focus is on the Minority-to-Majority transition. The  space parameters are  the proportions of respectively, the initial agents supporting opinion A $(p_0)$, the unavowed tie prejudices breaking in favor of opinion A, and the contrarians. The individual traits associated to $k$ and $x$ are invisible to agents. 

Previous studies have shown that ie breaking prejudice and contrarian were shown to have opposite effects \cite{min, cont}. On one hand, tie breaking prejudice produces the possibility of minority spreading by yielding two different asymmetric tipping points $p_T$ for opinion A and  $(1-p_T)$ for opinion B \cite{min}. On the other hand, contrarians reduce the gap between $p_T$ and  $(1-p_T)$ to eventually get them merged at fifty percent for rather low values of $x$ \cite{cont}.

But mixing both effects is found to lead to unexpected breaking of their respective impacts. In some range of proportions, contrarians are shown to favor the tie breaking prejudice as opposed to the precedent effect, where they restore a balance between both competing opinions. 

For every pair $(k, x)$, the dynamics in $p$ is found to be monitored by either a tipping point with two attractors in some cases, or predominantly by a single attractor. In the first case, opinion A (B) needs to gather a proportion of initial support larger than the tipping point to ensure a democratic victory over time. 

In the second more frequent case,  the outcome of the dynamics is unique and predetermined from the outset. Opinion A (B) cannot change the outcome, either victory or defeat depending on the location of the associated single attractor with respect to $50\%$. The outcome being independent of the initial supports  $p_0$ and  $(1-p_0)$.

The full results allow identifying the ranges of values of $(k, x)$ where an initial support $p_0<\frac{1}{2}$ ends up to $p_n>\frac{1}{2}$ after $n$ successive updates of individual opinions, thus turning democratically the minority to the majority without any deliberate and conscious manipulation, neither internal nor external.  The findings indicate that the expected democratic nature of free opinion dynamics is rarely met. Often, an initial minority of agents turn on their side a large fraction of agents initially supporting the opposite majority opinion.

In addition, the final outcome of a debate being predetermined, the only option for the supporters of the predetermined defeated  opinion, would be trying to modify the actual values of the pair $(k, x)$ to reach a location, which is beneficial to that  opinion. 

Nevertheless, identifying the means to implement changes in $k$ or and $x$ is out of the scope of the present paper. Moreover, such a strategy raises ethical issues about developing a model which could help manipulating public opinion. I address this issue  in the conclusion.

The rest of the paper is organised as follows: Section 2 reviews the spontaneous thwarting of democratic global balance in homogeneous populations for discussing groups of sizes $r=2, 3, 4$. Section 3 considers heterogenous agents with the introduction of  contrarians. Combined effects of contrarians and tie prejudice breaking is investigated in Section 4. The occurrence of a new unexpected regime of stationary alternating polarization is discussed in Section 5. The Conclusion contains a summary of the main results with a Word of caution about the responsibility of developing a model which could eventually lead to manipulate opinion dynamics.

\section{Spontaneous thwarting of democratic global balance in homogeneous populations}

Application of the Galam Majority Model (GMM) to opinion dynamics is based on iterating local majority rules to small groups of agents who are reshuffled after each update.  

For an homogeneous population of rational agents a democratically balance is obtained with a tipping point located at 50\% and two attractors at respectively 0\% and 100\%. Starting from a population divided in agents supporting two parties A and B with respective supports $p_0$ and $(1-p_0)$, a first cycle of local updates using discussing groups of size $r$ yields new supports $p_1$ and $(1-p_1)$.

Using odd size cells to guarantee the existence of a majority, $p_1>p_0$ if $p_0> 0.5$ and $p_1<p_0$ otherwise. A number $n$ of cycles is required to reach one of the two attractors. The value of $n$ is a function of $p_0$ and $r$, always less than 12 when $p_0<0.49$ and $p_0>0.51$.

However, a first thwarting of the global democratic balance occurs in case of even size groups. At a tie, assuming that silent breaking prejudices select A with probability $k$ and B with probability $(1-k)$, the 50\% tipping point splits in two tipping points located respectively at high and low values. 

One opinion can now turn majority spontaneously even starting a minority and subsequently an initial opposite majority shrinks to minority via repeated open mind discussions among small groups of rational agents. The update equations for group of size 2, 3 and 4 are,

\begin{equation}
p_{i+1,2,k}= p_{i,2,k}^2 + 2k p_{i,2,k} (1 - p_{i,2,k}) ,
\label{p12} 
\end{equation}

\begin{equation}
p_{i+1,3}= p_{i,3}^3 + 3 p_{i,3}^2 (1 - p_{i,3}) ,
\label{p13} 
\end{equation}

\begin{equation}
p_{i+1,4,k}= p_{i,4,k}^4 + 4p_{i,4,k}^3(1-p_{i,4,k})+6 k p_{i,4,k}^2 (1 - p_{i,4,k})^2 ,
\label{p14} 
\end{equation}
where $p_{i+1,r,k}$ ($p_{i+1,r}$) is the new proportion of A support after one cycle of updates from a proportion $p_{i,r,k}$ ($p_{i+1,r}$) using groups of size $r$, here with $r=2, 3, 4$. 

Odd sizes have no tie and are always independent of $k$. Accordingly, from here on, when the update is independent of $k$ or $k=\frac{1}{2}$ (balanced effect of prejudices), the parameter $k$ is not included in the indices defining $p$.

The three update equations yield the same two fixed points $p_B=0$ and $p_A=1$ with an additional tipping point for the last two, respectively $p_{T}=\frac{1}{2}$ and $p_{T,4,k}=\frac{1 - 6 k +\sqrt{13 - 36 k + 36 k^2}}{6 (1 - 2 k)}$, which in turn makes $p_B$ and $p_A$ attractors. For the first case ($r=2$), $k<\frac{1}{2}$ makes $p_B$ a tipping point and $p_A$ an attractor with the opposite for $k>\frac{1}{2}$. For  $k=\frac{1}{2}$ the update has no effect with $p_{i+1,2,k=1/2}=p_{i,2,k=1/2}$. In addition $p_{T,4,k=0}\approx 0.77$, $p_{T,4,k=1/2}= \frac{1}{2}$ and $p_{T,4,k=1}\approx 0.23$. Last case illustrates the phenomenon of minority spreading with any $p_0>0.23$ resulting in A winning. 

Above cases show how hidden prejudices thwart the democratic global balance of a dynamics obeying local majority rules in a homogeneous population of rational agents. Only the case $r=3$ insures a democratic balance due to the absence of ties in the discussing groups. 

\begin{figure}[ht]
\hspace{-0.5cm}
\includegraphics[width=0.5\textwidth]{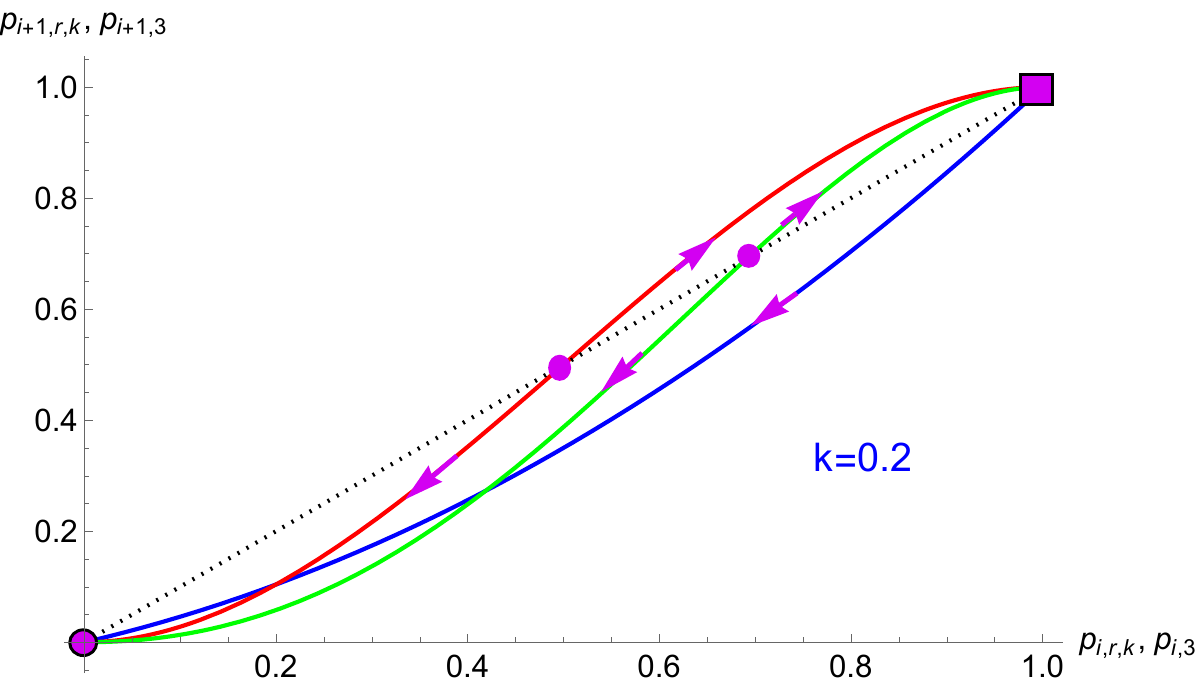} \quad
\includegraphics[width=0.5\textwidth]{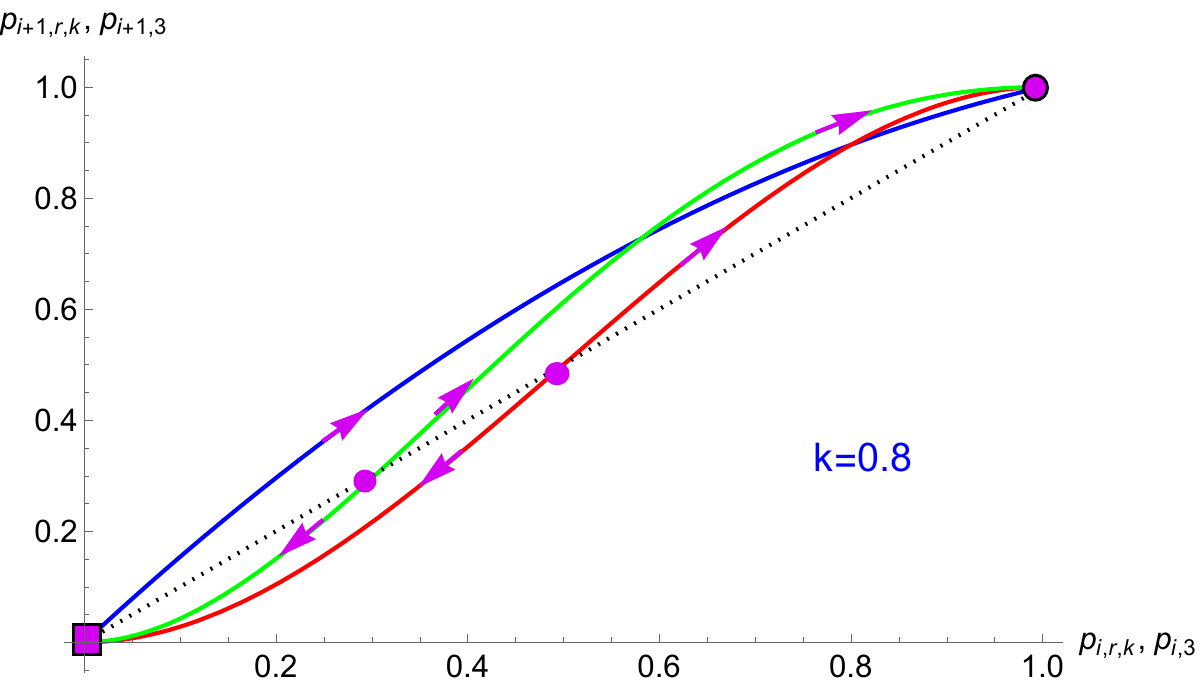}
\caption{Update curves $p_{i+1,r,k}$ as a function of $p_{i,r,k}$ for $r=2$ (blue curves) and $r=4$ (green curves) at $k=0.2$ (left) and $k=0.80$ (right). Update curve for 
$p_{i+1,3}$ (r=3) as a function of $p_{i,3}$ is also shown (red curves). The dotted lines represent the diagonal $p_{i+1}=p_{i}$. Arrows indicate update directions, solid circles indicate tipping points, circled solid circles indicate attractors and squares indicate either attractors or tipping points depending on $r$ and $k$. Only $r=3$ yields a democratic balance with $p_{T}=\frac{1}{2}$.}
\label{h12}
\end{figure}

Figure (\ref{h12}) shows the update curves $p_{i+1,r,k}$ as a function of $p_{i,r,k}$ for $r=2$ and $r=4$ at $k=0.2$ and $k=0.80$. The curve $p_{i+1,r=3}$ as a function of $p_{i,r=3}$ is also shown. The directions of the update flows are also indicated with the tipping points and atractors.

\section{Heterogeneous agents: the contrarian thwarting} 

I now introduce a proportion $x$ of contrarian agents with a proportion $(1-x)$ of rational agents. A contrarian agent is identical to a rational agent beside the fact that once the discussion is over, they do no follow the majority opinion but shift to the opposite whatever is the majority opinion. Contrarians  are not identifiable. The associated update equation for groups of size $r$ is, 
\begin{eqnarray}
p_{i+1,r,x}&=& (1-x) p_{i+1,r,x} + x  \left\{ 1-p_{i+1,r,x}  \right\} ,  \nonumber \\
&=& (1-2x) p_{i+1,r,x}  + x  ,
\label{p1rx} 
\end{eqnarray}
where $p_{i+1,r,x}$ has been defined above and without prejudice effect with either odd sizes or $k=\frac{1}{2}$ for even sizes.

\subsection{Size 2} 

Starting to study the contrarian impact on the landscape of a democratic dynamics of opinion with discussing groups of size 2, the update equation Eq. (\ref{p1rx}) becomes,
\begin{equation}
p_{i+1,2,x}= (1-2x)  p_{i,2,x} + x  ,
\label{p12x} 
\end{equation}  
with the unique attractor $p_{A,B,2}=\frac{1}{2}$ provided $x\neq 0$. At $x=0$, $p_{i+1,2,x=0}=p_{i,2,x=0} .$ For whatever initial respective supports for A and B, even a handful of contrarians drives the dynamics towards  fifty-fifty creating a hung election outcome, which is non democratic. Moreover, any actual measure does not yield the expected fifty-fifty due to incompressible statistical fluctuations but an outcome very close to $0.50$. Accordingly, one opinion does win with a "wrong". The winner is thus de facto the result of chance, which leads the loser to question the actual winner as being the result of fraud. An exact counting would yield no winner with 50\% for each choices.

\subsection{Size 3}

For groups of size 3 the associated update equation writes,
\begin{equation}
p_{i+1,3,x}=(1-2x)\left\{p_{i,3,x}^3 + 3 p_ {i,3,x}^2 (1 - p_{i,3,x}) \right\}    + x  ,
\label{p13x} 
\end{equation}
whose dynamics is driven by the two attractors,
\begin{equation}
p_{B(A),3,x}=\frac{1 - 2 x \mp \sqrt{1 - 8 x + 12 x^2}}{  2 (1 - 2 x)} ,
\label{p3BAx} 
\end{equation}
and the tipping point $p_{T}=\frac{1}{2}$.

\begin{figure}
\hspace{-0.8 cm}
\includegraphics[width=0.5\textwidth]{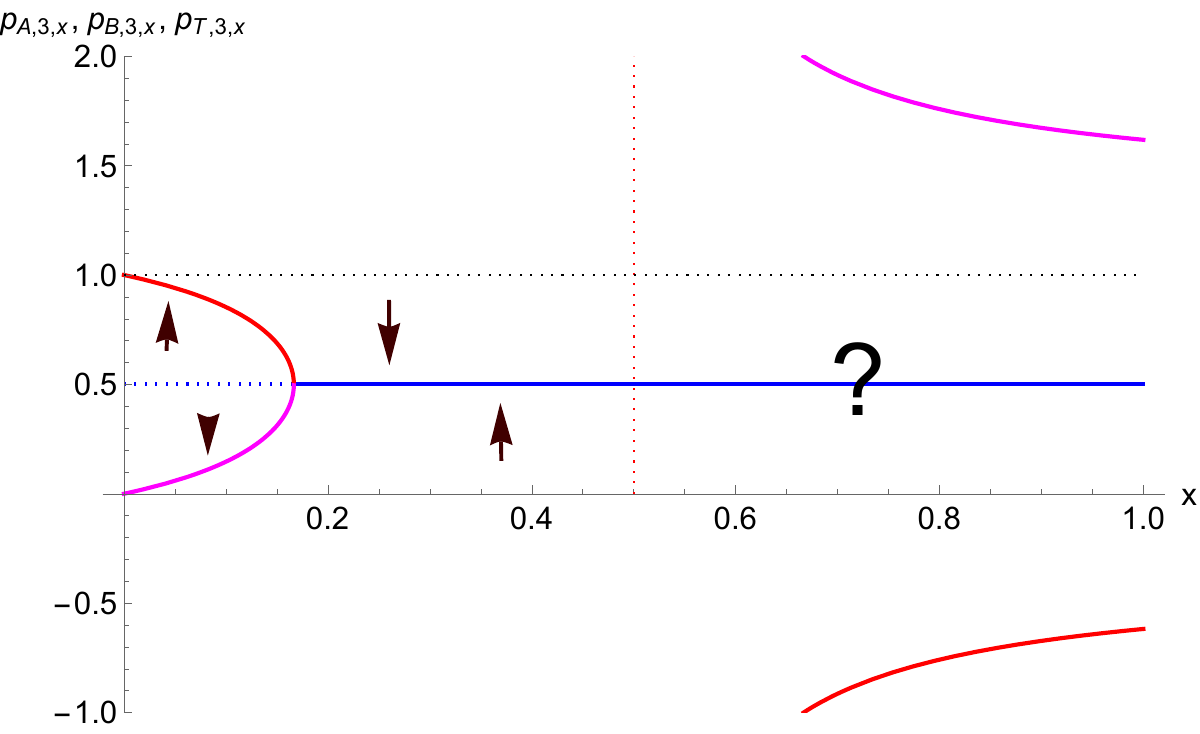} \quad
\includegraphics[width=0.5\textwidth]{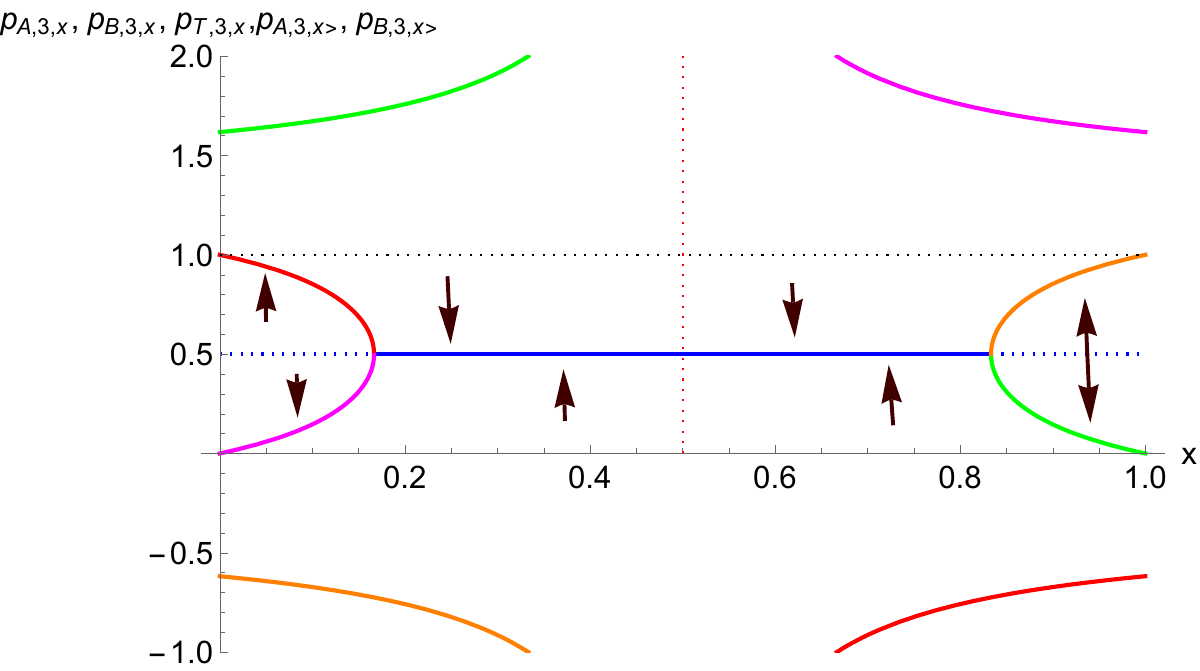}

\vspace{0.5cm}
\centering
\includegraphics[width=0.8\textwidth]{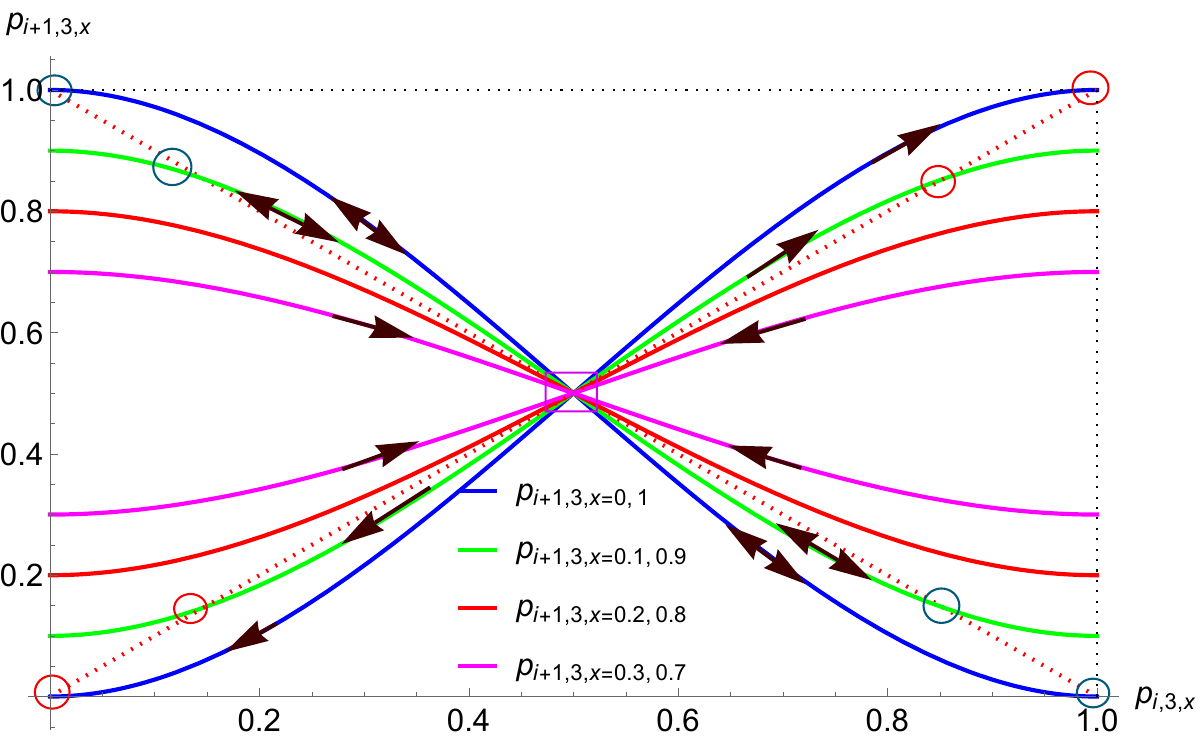}

\caption{Attractors and tipping points yielded by Eq. (\ref{p13x}) are shown on the upper left side for $r=3$ as a function of the proportion $x$ of contrarians. Only the values between 0 and 1 are meaningful. Solid lines (in red, magenta, blue) are attractors while tipping points (in dotted bliue) exists only for $x<\frac{1}{6}\approx 0.17$. When $x>\frac{1}{6}$ one unique attractor drives the dynamics toward perfect equalities. For  $x>\frac{1}{2}$ more than half the population being contrarians, alternating dynamics is expected. The upper right side shows the completed dynamics with a single attractor dynamics for $\frac{1}{6}< x<\frac{5}{6}$ and a dynamics with two alternating attractors when $x>\frac{5}{6}$. Update curves $p_{i+1,3,x}$ as a function of $p_{i,3,x}$ for $x=0, 0.10, 0.20,0.30$ and  $x=0.70, 0.80, 0.90, 1$ are shown in the lower part. The dotted red lines represent the diagonals $p_{i+1}=p_{i}$ and   $p_{i+1}=1-p_{i}$. Arrows indicate update directions, red circles indicate attractors and blue circles indicate alternating attractors. The $p_{T}=\frac{1}{2}$ square in the middle is either a tipping point when $x<\frac{1}{6}$ and $x>\frac{5}{6}$ or an attractor when $\frac{1}{6}< x<\frac{5}{6}$. The lower part shows $p_{i+1,3,x}$ as a function of $p_{i,3,x}$ for respectively $x=0, 0.10, 0.20,0.30, 0.70, 0.80, 0.90, 1$. Arrows indicate the direction of the updates. Double arrows signal an alternating dynamics.}
\label{c12}
\end{figure}

However, $p_{B,3,x}$ and $p_{A,3,x}$ are defined only as long as $1 - 8 x + 12 x^2\geq 0$ and $0\leq p_{B,3,x}, p_{A,3,x} \leq1$, which is satisfied in the range $x\leq \frac{1}{6}\approx 0.17$. At this stage the dynamics is globally democratic with the spreading of the initial majority. The contrarian effect prevents reaching unanimity.
In addition, at $x=\frac{1}{6}$ the two attractors merge at the tipping point turning it to the unique attractor of the dynamics. When $x\geq \frac{1}{6}$ above democratic dynamics is thus suddenly puts upside down with a single attractor dynamics at $p_{T}=\frac{1}{2}$. Any initial conditions end up at 50\% breaking the previous democratic balance. These results are exhibited in the left side of Figure (\ref{c12}).

The update curves $p_{i+1,3,x}$ as a function of $p_{i,3,x}$ for $x=0, 0.10, 0.20,0.30$ are shown in the lower part of Figure (\ref{c12}). The directions of the update flows are also indicated with the tipping points and attractors.

\begin{figure}
\hspace{-0.5cm}
\includegraphics[width=0.5\textwidth]{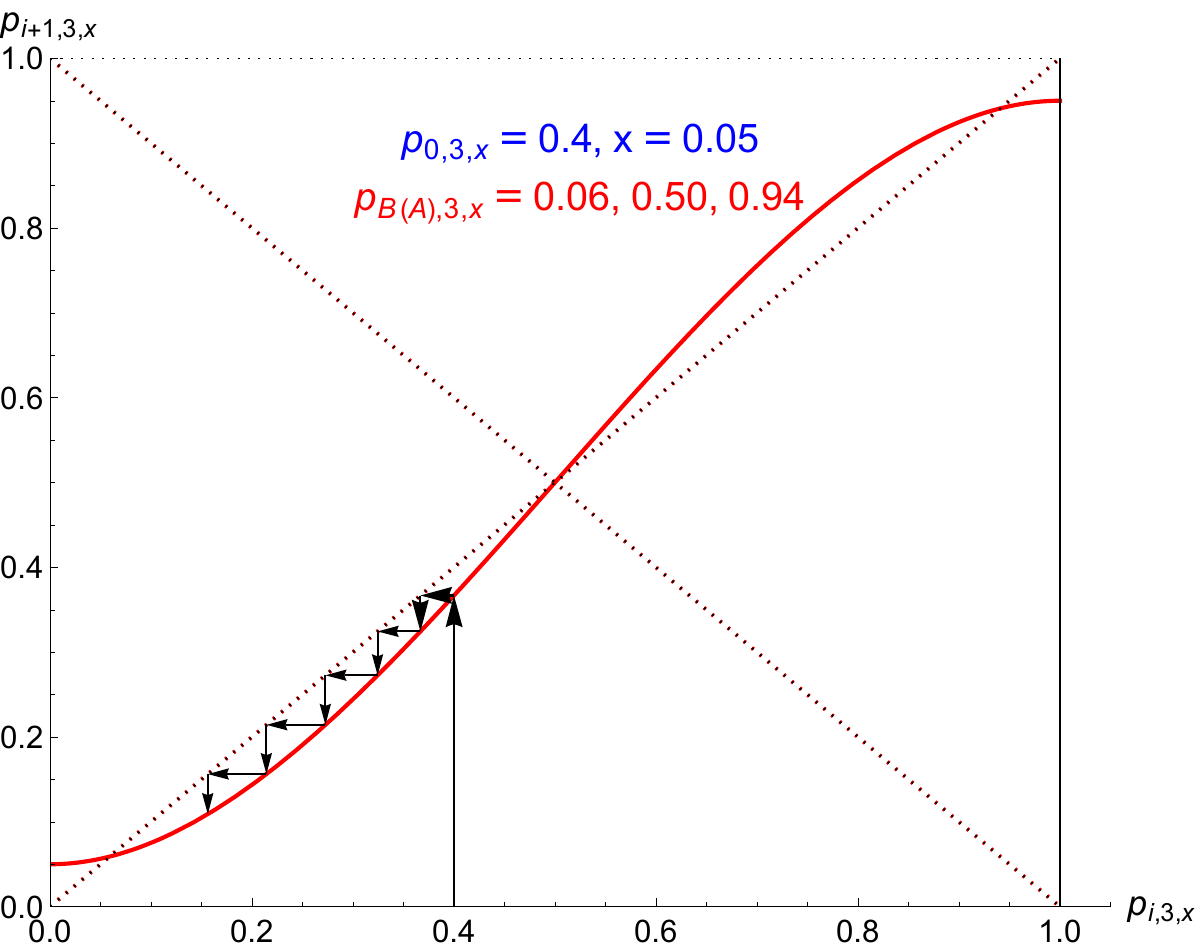} \quad
\includegraphics[width=0.5\textwidth]{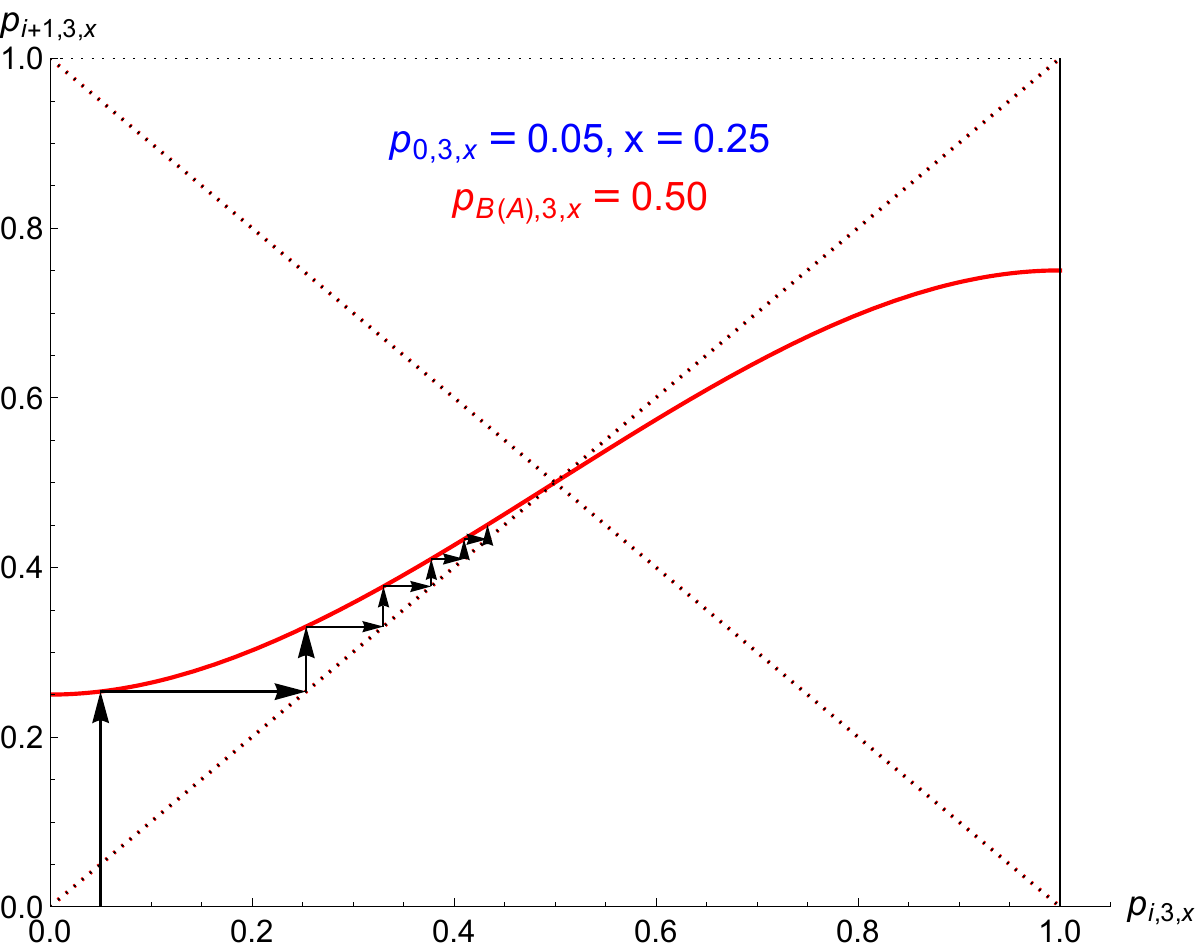}

\vspace{0.5cm}
\hspace{-0.5cm}
\includegraphics[width=0.5\textwidth]{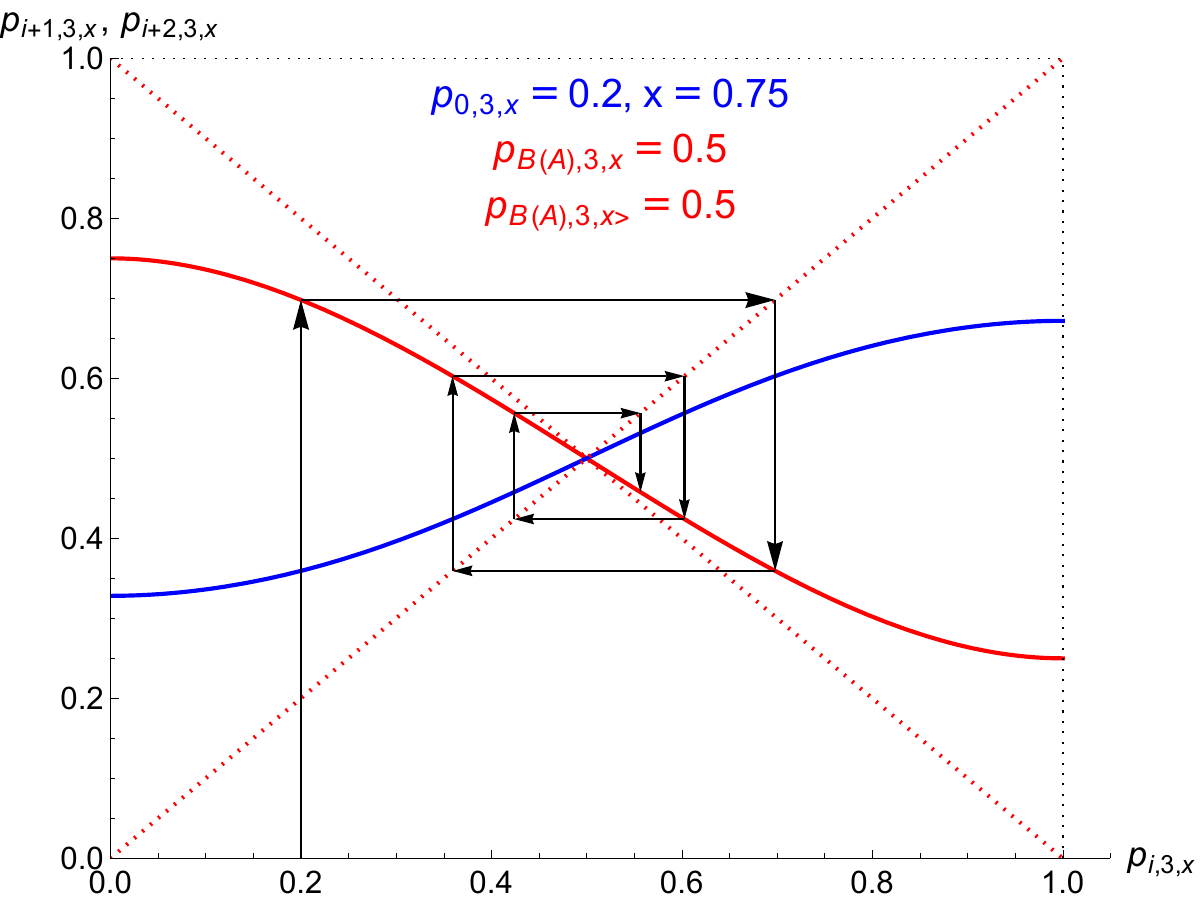} \quad
\includegraphics[width=0.5\textwidth]{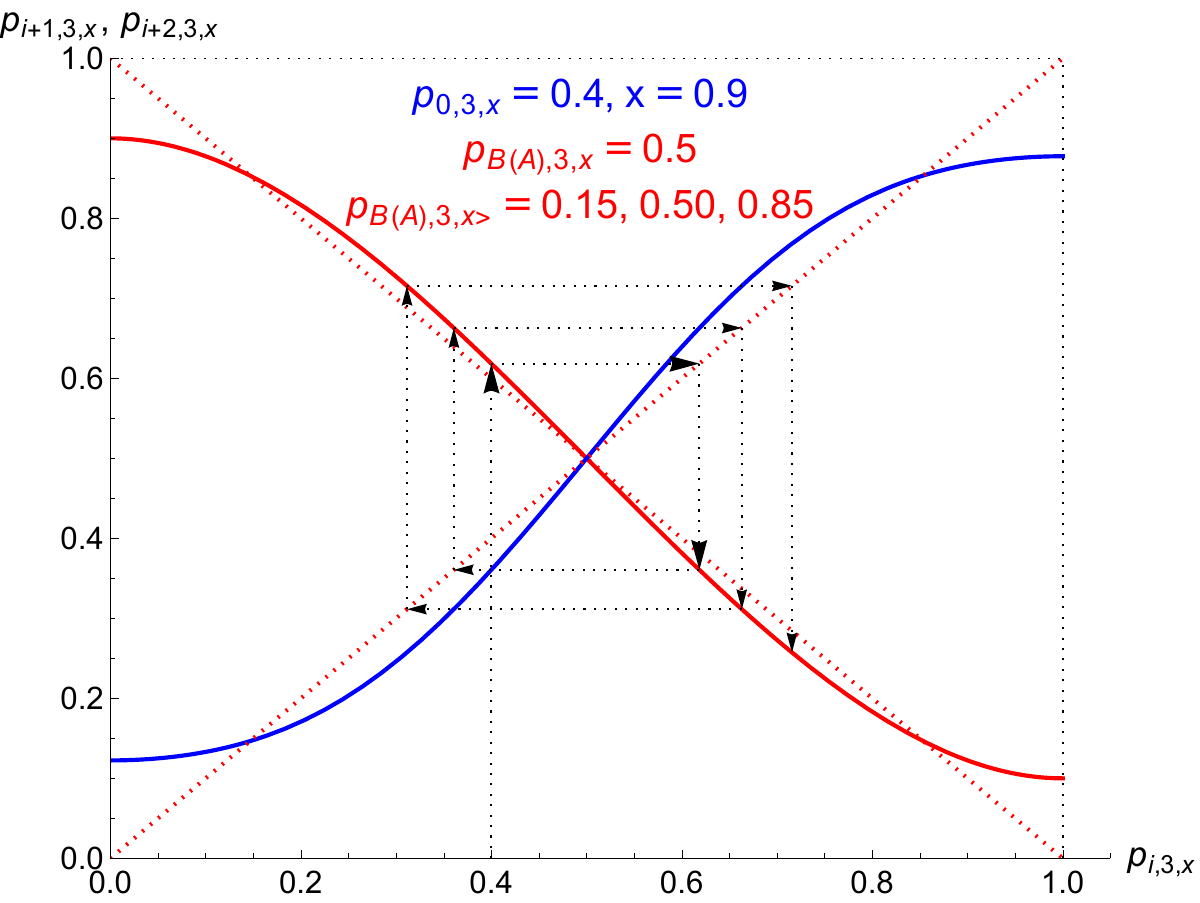}
\caption{The upper left part shows the update Eq. (\ref{p13x}) for $x=0.105$, which has a tipping point at 0.50  with two attractors at respectively 0.06 and 0.94. The dynamics is illustrated from $p_{0,3,x}=0.40$. The upper right part shows the case $x=0.25$, which is a single attractor dynamics with one attractor at 0.50. The dynamics is illustrated from $p_{0,3,x}=0.05$. The lower part exhibits both updates  $p_{i+1,3,x}$ in red and  $p_{i+2,3,x}$ in blue as function of  $p_{i,3,x}$. The left side shows the case $x=0.75$, which is an alternating single attractor dynamics with one attractor at 0.50. The dynamics is illustrated from $p_{0,3,x}=0.20$. The right side shows the case $x=0.90$, which is an alternating dynamics with a tipping point at 0.50 and two alternating attractors at respectively 0.15 and 0.85. The dynamics is illustrated from $p_{0,3,x}=0.40$.}
\label{cc3456}
\end{figure}

In addition, I notice that in the range $\frac{1}{2} < x \leq 1$ the dynamics is peculiar with more than half a community being contrarian. Although such a situation could sound socially awkward it is interesting to investigate the related dynamics. The upper right part of Figure (\ref{c12}) and the low part of Figure (\ref{cc3456}) exhibit a symmetry between ranges $x<\frac{1}{2}$ and $x>\frac{1}{2}$ with an alternating update performed for the second case. This effect is also observed in the lower part of Figure (\ref{c12}) between the update curves for respectively $x=0, 0.10, 0.20,0.30$ and  $x=0.70, 0.80, 0.90, 1$.

This statement is proven solving the Equation $p_{i+2,3,x}=p_{i,3,x}$ to determine all its nine solutions. In addition to the above two attractors $p_{B(A),3,x}$ and tipping point $p_{T}=\frac{1}{2}$, four solutions are found to be complex and last two write,
\begin{equation}
p_{B(A),3,x>}=\frac{1 - 2 x \pm \sqrt{5 - 16 x + 12 x^2}}{  2 (1 - 2 x)} ,
\label{p3BAx>} 
\end{equation}
which become identical to $p_{B(A),3,x}$ by substituting $(1-x)$ to $x$. Similarly to $p_{B(A),3,x}$ valid in the range $0 \leq x \leq 1/6$ the attractors $p_{B(A),3,x>}$ are valid only in the range $5/6 \leq x \leq 1$.

Therefore, the function $p_{i+2,3,x}$ exhibits a single attractor dynamics located at $\frac{1}{2}$ as a function of $p_{i,3,x}$ for  $1/2 \leq x \leq 5/6$ as seen from the lower left part of Figure (\ref{cc3456}). The lower right part shows a threshold dynamics as expected for $5/6 \leq x \leq 1$. It is worth noticing that in this case the associated dynamics is alternating between the two attractors $p_{B(A),3,x>}$ for $p_{i+1,3,x}$ as a function of $p_{i,3,x}$. They are alternating attractors. These attractors can be also obtained solving $p_{i+1,3,x}=1-p_{i,3,x}$ due to the symmetry of the update Eq. (\ref{p13x}).

\subsection{Size 4} 

Along size 2 to study the impact of contrarian agents putting $k=\frac{1}{2}$ avoids a possible confusion with the prejudice effect. The update equation Eq. (\ref{p1rx}) then writes,
\begin{equation}
p_{i+1,4,x}=(1-2x)\left\{p_{i,4,x}^4 + 4 p_{i,4,x}^3 (1 - p_{i,4,x})+3 p_{i,4,x}^2 (1 - p_{i,4,x})^2 \right\} + x  ,
\label{p14x} 
\end{equation}
which is found to be identical to Eq. (\ref{p13x}) provided $k=\frac{1}{2}$.

The dynamics is thus driven by the same landscape as for $r=3$ as above in Figures (\ref{c12}, \ref{cc3456}).

\section{Combined thwarting effect of prejudices and contrarians}

The spontaneous thwarting created by prejudice tie breaking at even sizes favors one choice over the other. In contrast, the contrarian effect tends to first smooth and then suppress any difference between the two choices. Yet, both effects break the democratic balance associated to the aggregated initial majority. 

I now investigate the combined effect of simultaneous prejudice breaking and contrarians for the two cases $r=2$ and $r=4$. The case  $r=3$ is not considered in this Section since no prejudice effect occurs for odd sizes.

\subsection{Size 2} 

When prejudice breaking is added to contrarians, the update Equation Eq. (\ref{p12x}) becomes,
\begin{equation}
p_{i+1,2,k,x}= (1-2x)  \left\{  p_{i,2,k,x}^2 + 2k p_{i,2,k,x} (1 - p_{i,2,k,x})  \right\}  + x  ,
\label{p12kx} 
\end{equation}  
whose fixed points are given by,
\begin{equation}
p_{B(A),2,k,x}=\frac{1 - 2 k+4 k x \mp \sqrt{(1 - 2 k+4 k x)^2- 4 x (1- 2 k) (1 - 2 x)}}{  2  (1- 2 k) (1 - 2 x)} ,
\label{p2ABkx} 
\end{equation}
with the radical in the square root being always positive for $0\leq k \leq1$ and $0\leq x \leq1$. 

\begin{figure}
\hspace{-0.5cm}
\includegraphics[width=0.5\textwidth]{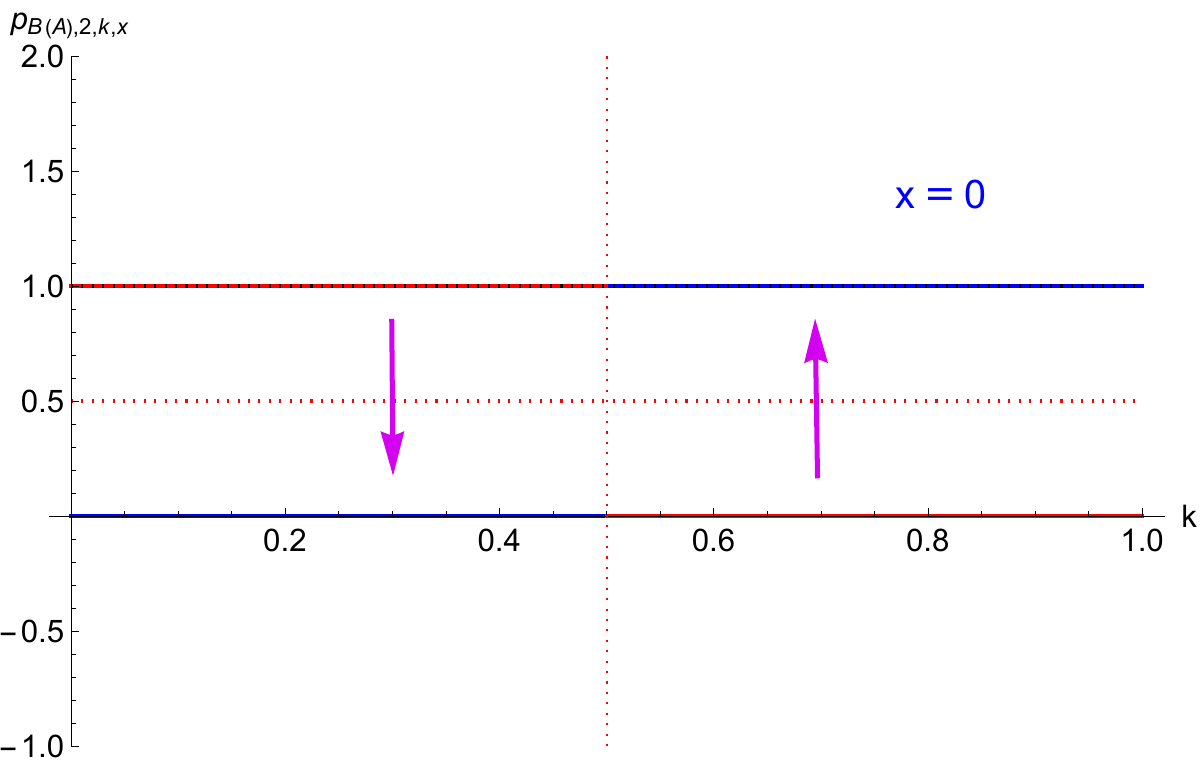} \quad
\includegraphics[width=0.5\textwidth]{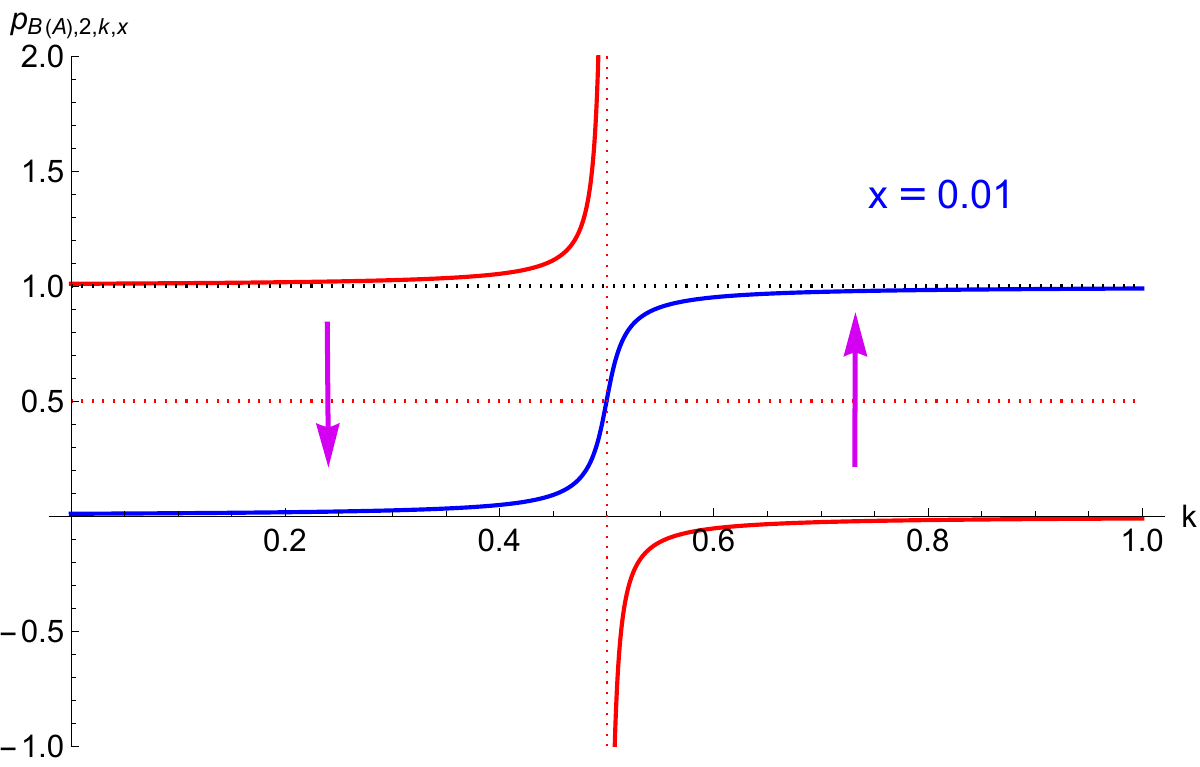} 

\vspace{0.5cm}
\hspace{-0.5cm}
\includegraphics[width=0.5\textwidth]{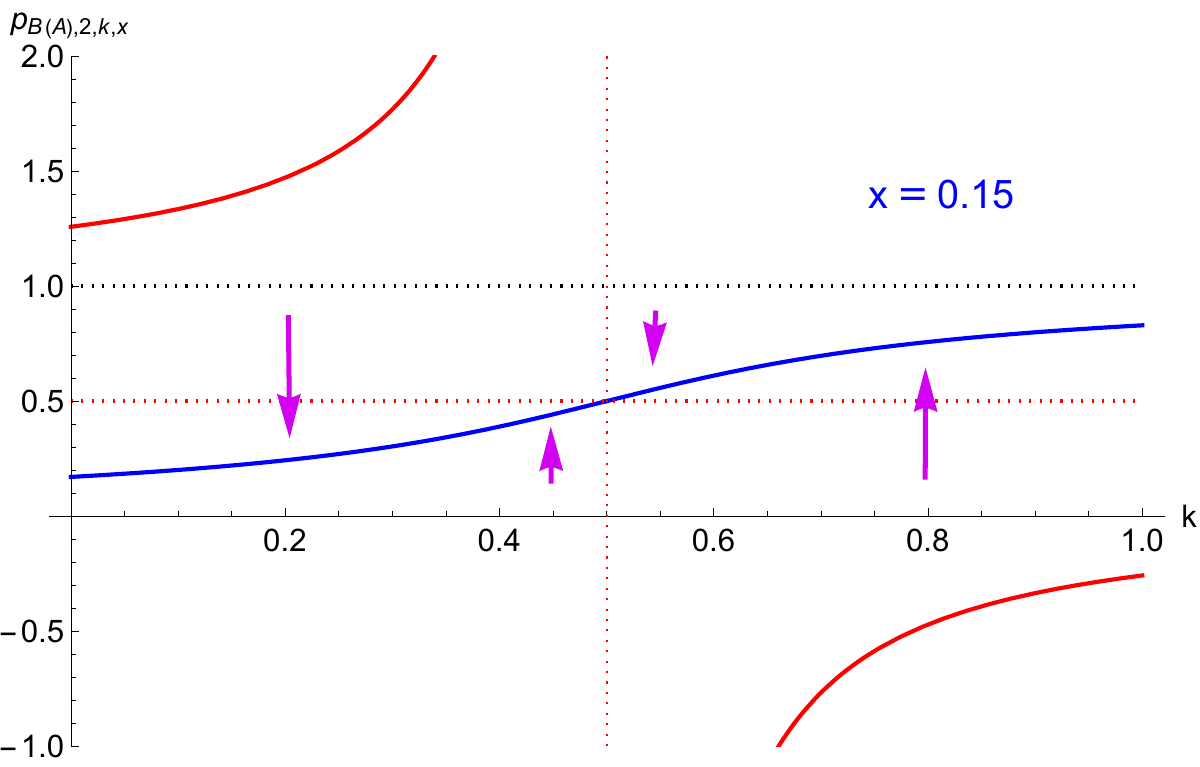} \quad
\includegraphics[width=0.5\textwidth]{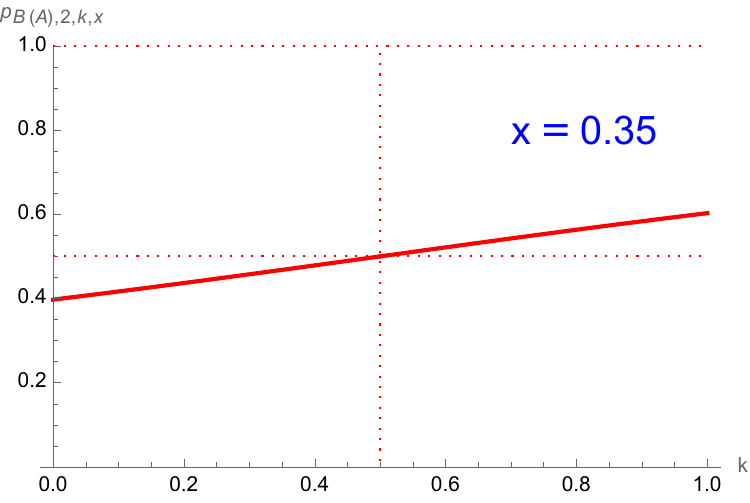} 

\vspace{0.5cm}
\hspace{-0.5cm}
\includegraphics[width=0.5\textwidth]{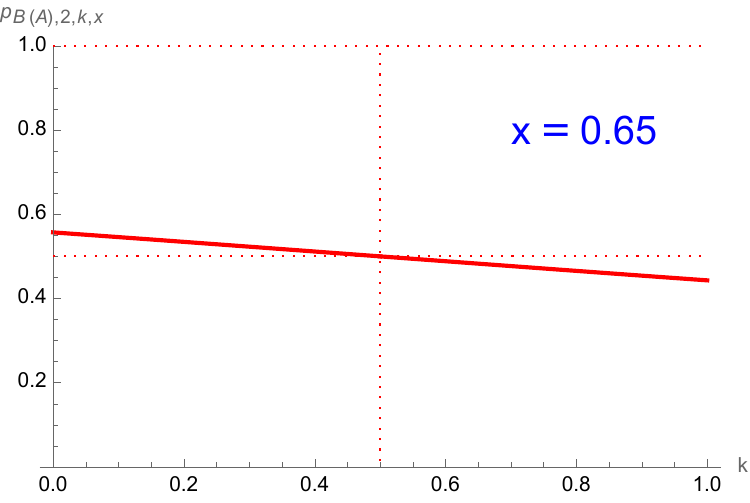} \quad
\includegraphics[width=0.5\textwidth]{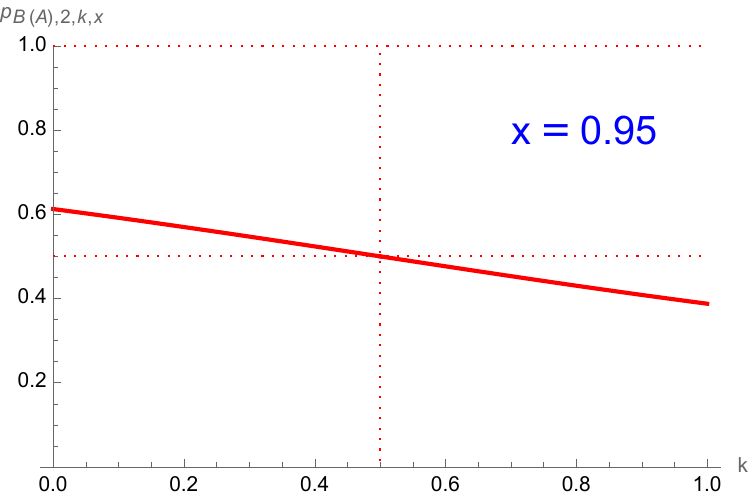}
\caption{The upper left part of the Figure shows the attractors and tipping points of the dynamics as a function of $k$ when $x=0$ for groups of size 2. The upper right part shows the effect of a tiny proportion of contrarians $(x=0.01)$ turning the dynamics to single attractor. The prejudice effect is simultaneously slightly reduced. A larger proportion of contrarian accentuates the reduction effect of prejudices as seen in the middle left part with $x=0.15$. From $x\approx 0.30$ till $x=1$, $p_{B,2,k,x}$ is quasi-linear as a function of $k$ for a given $x$  as illustrated in the middle right part and left and right lower part for respectively $x=0.35, 0.65, 0.95$.}
\label{2kxa}
\end{figure}

At $x=0$ the values of Section 2 are recovered with $p_{B(A),2,k,x=0}=0(1)$. However, as soon as $x\neq 0$ the fixed point $p_{A,2,k,x}$ is no longer valid being outside the interval $[0,1]$ as seen in Figure (\ref{2kxa}). In the range $]0,1]$ the unique valid fixed point is thus $p_{B,2,k,x}$.  It is located lower than $\frac{1}{2}$ in the range $0\leq k<\frac{1}{2}$ and higher than $\frac{1}{2}$ for $\frac{1}{2}<k\leq 1$. At  $k=\frac{1}{2}$, $p_{B,2,k,x=1/2}=\frac{1}{2}$.

Moreover, $p_{B,2,k,x}$ is quasi-linear as a function of $k$ for a given $x$ from $x\approx 0.30$ till $x=1$  as illustrated in Figure (\ref{2kxa}). In addition, for $\frac{1}{2} \leq x \leq 1$, no much change occurs with a slight variation of the associated slope. In the range $0< x< \frac{1}{2}$, $p_{B,2,k,x}$ is the attractor of the dynamics. 

\subsubsection{Dynamics $x>\frac{1}{2}$: part 1} 

For $x>\frac{1}{2}$, contrarians being majority turns the dynamics oscillating. In this range, the case $r=3$ has revealed an alternating tipping point regime for $x>\frac{5}{6}$. This regime could be anticipated by symmetry from the tipping point regime observed for $x<\frac{1}{6}$. In contrast, here for $r=2$, there is no tipping point regime for low concentration of contrarians $x< \frac{1}{6}$. In addition there is no symmetry between $x$ and $1-x$ as for $r=3$. 

These two facts hint against the existence of an alternating tipping point regime at high concentration of contrarians. However, solving the fixed point equation $p_{i+1,2,k,x}=p_{i,2,k,x}$ of the double iteration reveals an unexpected behavior at very high concentration of contrarians. The fixed point equation being a polynomial of degree 4, two new fixed points,
\begin{equation}
p_{B(A),2,k,x>}=\frac{-1 - 2 k+4 k x \mp \sqrt{ -3-4k(1-k)(1-2x)^2-4x(1-2x)}}{2  (1- 2 k) (1 - 2 x)} ,
\label{p2^2ABkx>} 
\end{equation}
are obtained in addition to $p_{B(A),2,k,x}$ also given by Eq. (\ref{p2ABkx}). To be valid the additional fixed points must obey $0 \leq p_{B(A),2,k,x>} \leq 1$ with  $ -3-4k(1-k)(1-2x)^2-4x(1-2x) \geq 0$, which happens for,
\begin{equation}
x>\frac{1 - 4 k + 4 k^2 + \sqrt{7 - 12 k + 12 k^2}}{4 (1 - 2 k + 2 k^2)}  ,
\label{2xk>} 
\end{equation}
where $7 - 12 k + 12 k^2 \geq 4$ for $0\leq k\leq 1$.

\begin{figure}
\hspace{-0.5cm}
\includegraphics[width=0.5\textwidth]{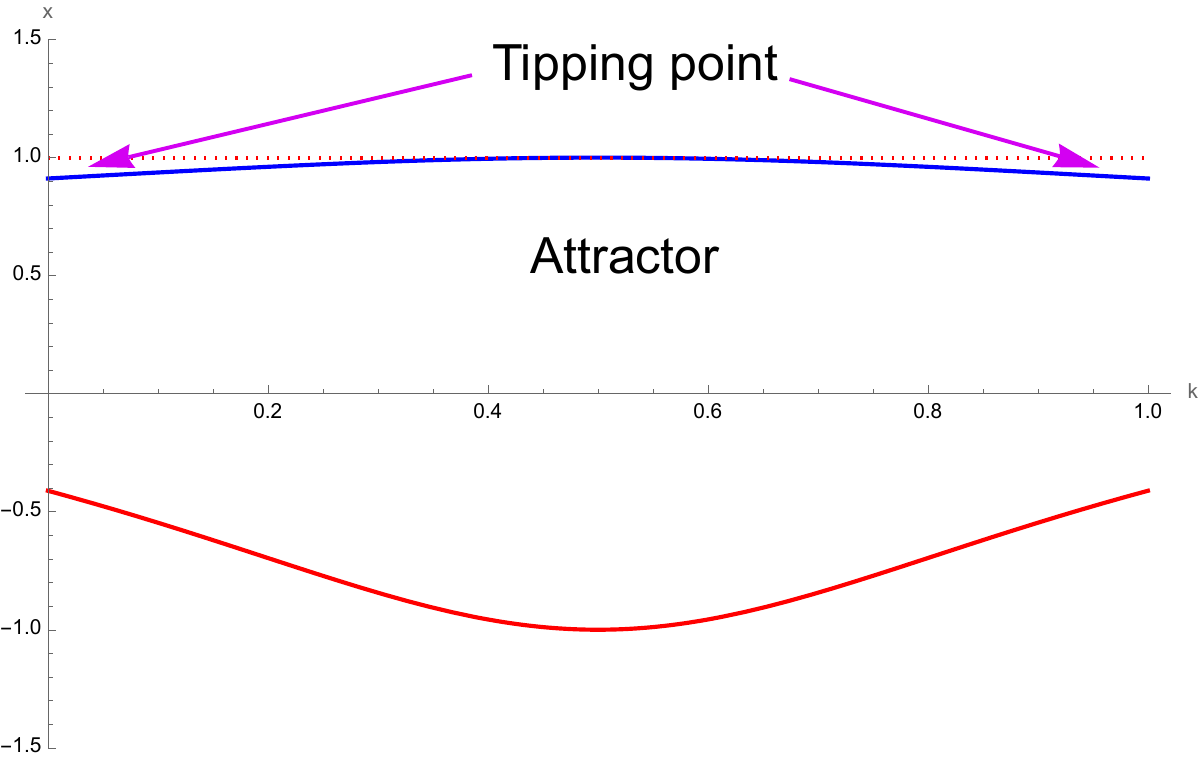} \quad
\includegraphics[width=0.5\textwidth]{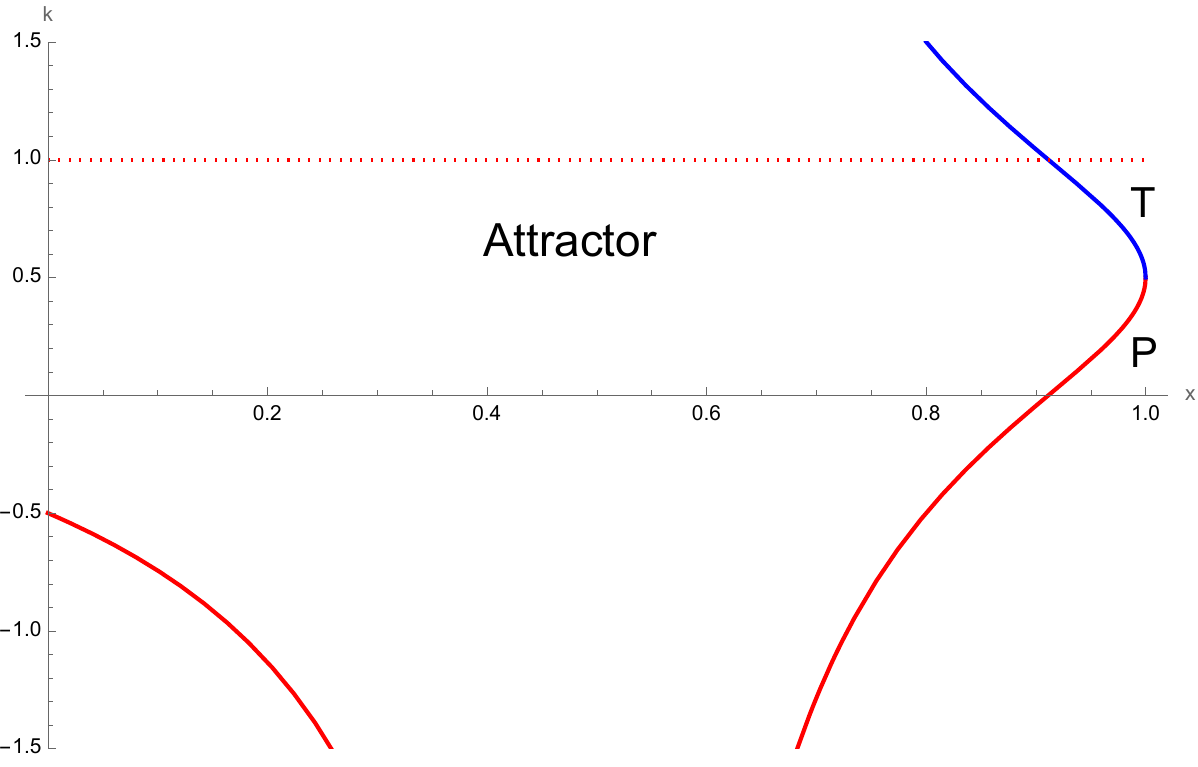}

\vspace{0.5cm}
\hspace{-0.5cm}
\includegraphics[width=0.5\textwidth]{2kx6.pdf}\quad
\includegraphics[width=0.5\textwidth]{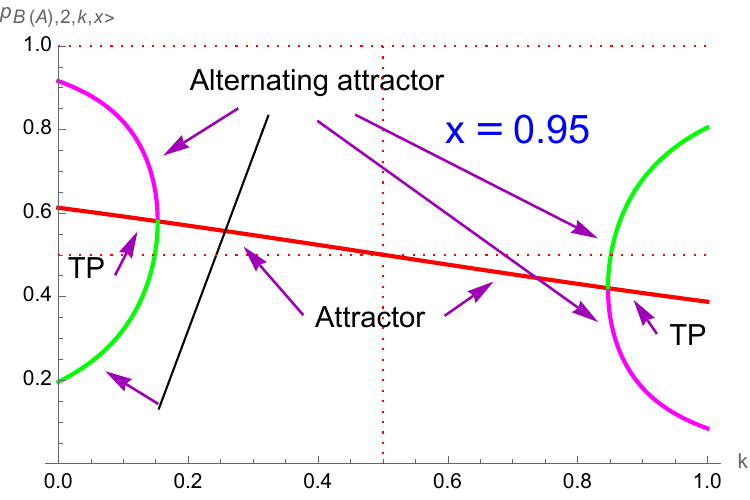}

\vspace{0.5cm}
\hspace{-1cm}
\includegraphics[width=0.55\textwidth]{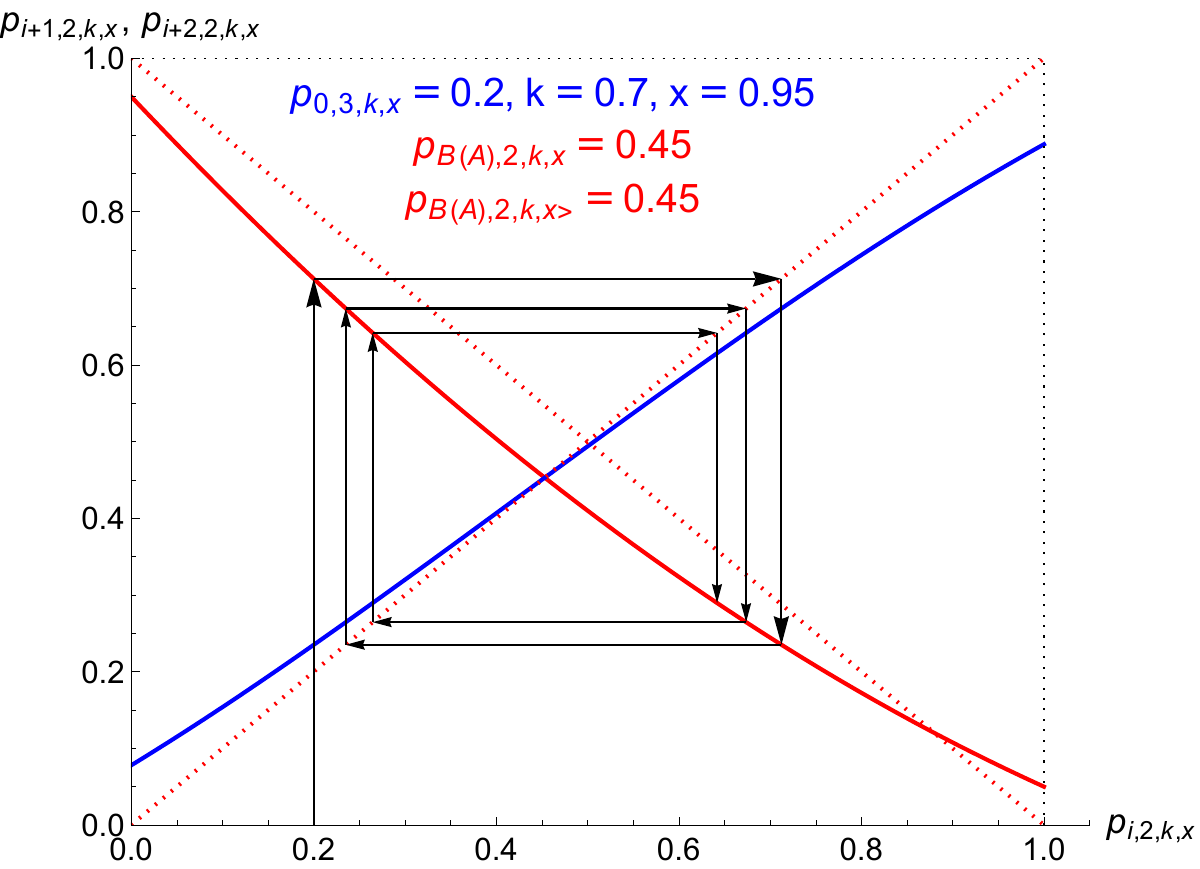} \quad
\includegraphics[width=0.55\textwidth]{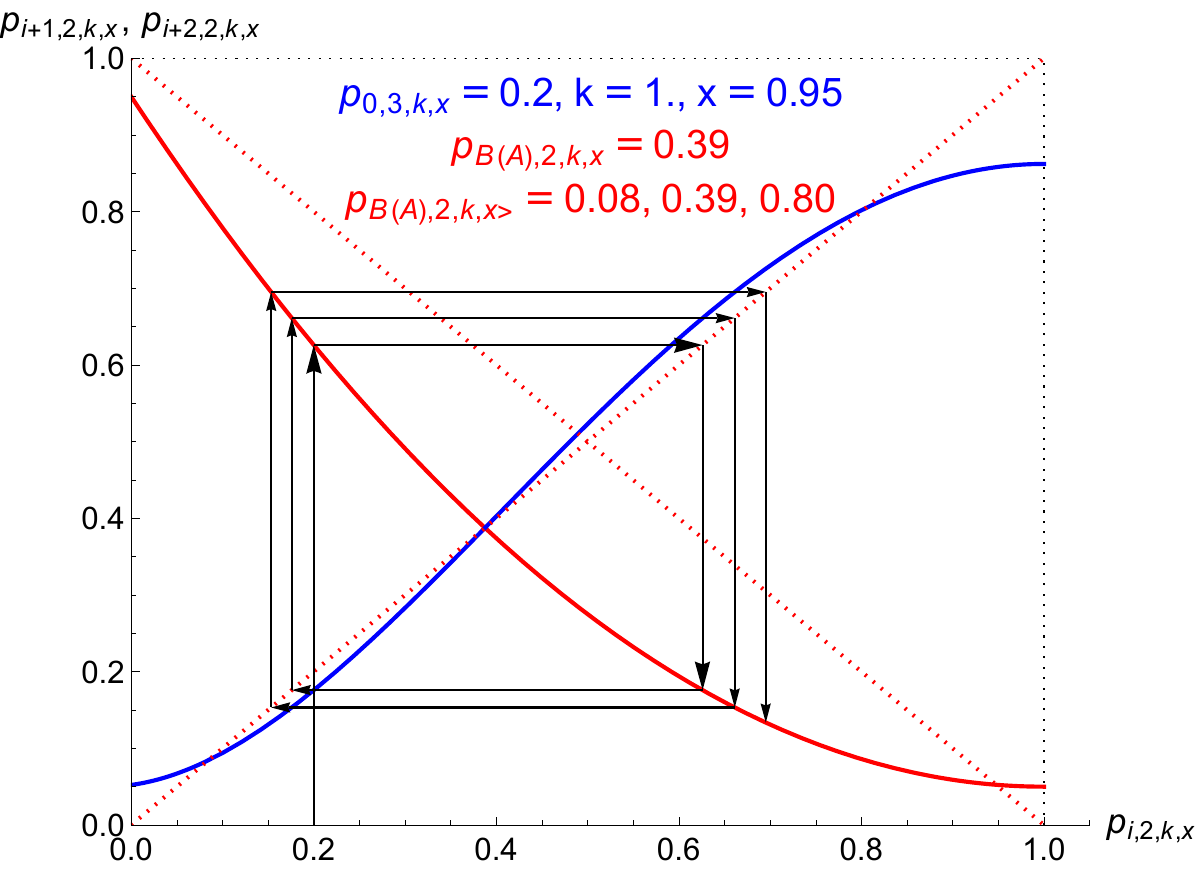} 
\caption{The upper part of the Figure shows the domains of respectively a single attractor dynamics and an alternating tipping point dynamics for the update $p_{i+1,2,k,x}$. Left part shows them for $x$ as a function of $k$ and the right part for $k$ as a function of $x$ (TP = tipping point). The middle part shows the single fixed point $p_{B,2,k,x}$ for $x=0.95$ as a function of $k$. On the left part when only the update $p_{i+1,2,k,x}$ is used. On the right part when the stability of the fixed point has been added using either $p_{i+2,2,k,x}$ or the derivative of $p_{i+1,2,k,x}$ with respect to $p_{i,2,k,x}$ at the fixed point. The lower part exhibits both $p_{i+1,2,k,x}$ and $p_{i+2,2,k,x}$ as a function of $p_{i+1,2,k,x}$ for $x=0.95$ with $k=0.7$ on the left and $k=1$ on the right. Arrows shows the evolution of $p_{i,2,k,x}=0.20$ for six successive updates for each case.}
\label{2kxb}
\end{figure}

Figure (\ref{2kxb}) shows that $p_{B(A),2,k,x>}$ exists only in a narrow range of very high values of $x$ as a function of $k$ with at minimum $x=\frac{1+ \sqrt{7}}{4} \approx 0.91$ at $k=0$ and $k=1$ with $x=1$ at $k=\frac{1}{2}$.

The domain of validity of the additional fixed points can also be defined for $k$ as a function of $x$ instead of $x$ as a function of $k$ given by Eq. (\ref{2xk>}).  The related condition writes,
\begin{equation}
k<\frac{1}{2}-\frac{ \sqrt{1-x^2}}{(1 - 2x)}  \;  \;  \lor \;  \;  k>  \frac{1}{2}+\frac{ \sqrt{1-x^2}}{(1 - 2x)}  ,
\label{2kx>} 
\end{equation}
whose associated domain is also shown in Figure (\ref{2kxb}). The domain extension is also very narrow as in the case of $x$ as function of $k$ with 

Implementing above findings allows to enrich the landscape diagram of the dynamics for $x>\frac{1}{2}$ as exhibited in the middle part of Figure (\ref{2kxb}).  For every pair $k$ and $(1-k)$ there exists one range of $x>\frac{1+ \sqrt{7}}{4}$ for which an alternating tipping point regime is activated as seen in the left upper part of Figure (\ref{2kxb}). But for most values of $x$ no alternating tipping point regime takes place with instead one single attractor dynamics. Only in the range $x>\frac{1+ \sqrt{7}}{4}$ two ranges of $k$ respectively at low and large values exist as given by Eq. (\ref{2kx>}) for which an alternating tipping point dynamics is created as seen in the right upper part of Figure (\ref{2kxb}).

\subsubsection{Dynamics $x>\frac{1}{2}$: part 2} 

Alternatively, I can skip analyzing the double update equation and instead investigate directly the nature of the unique fixed point $p_{B,2,k,x}$. If $p_{B,2,k,x}$ is stable,  it is an attractor, whereas if it is unstable, it is a tipping point. Each regime is then determined by the value of the derivative of $p_{i+1,2,k,x}$ with respect to  $p_{i,2,k,x}$ taken at the fixed point, which yields,
\begin{equation}
\lambda_{2,k,x} \equiv \frac{\partial p_{i+1,2,k,x}}{\partial p_{i,2,k,x}} \Bigg|_{p_{B,2,k,x}} = 1 - \sqrt{ (1 -2 k (1 -2 x))^2-4 x(1- 2 k)  (1- 2 x) } ,
\label{lam} 
\end{equation}  
using Eqs.(\ref{p12kx}) and (\ref{p2ABkx}). The condition $-1<\lambda_{2,k,x} <1$  implies the stability of $p_{B,2,k,x}$, which is then an attractor. In contrast, $\lambda_{2,k,x} <-1$ or $\lambda_{2,k,x} >1$ makes $p_{B,2,k,x}$ unstable, thus becoming a tipping point. The domain associated with the tipping regime is determined by,
\begin{equation}
0 \leq k <  \frac{1}{2} \;  \;  \land \;  \;    \frac{1}{2}  <k \leq 1 \;  \;  \land \;  \;   \frac{1 - 4 k + 4 k^2 + \sqrt{7 - 12 k + 12 k^2}}{4 (1 - 2 k + 2 k^2)} <x \leq 1 ,
\label{lam-k} 
\end{equation} 
 with $\lambda_{2,k,x} =1$ for  $k = \frac{1}{2}$. In this case $p_{B,2,k,x}$ is an attractor. These conditions can be recast as,
\begin{equation}
\frac{1+ \sqrt{7}}{4} < x \leq 1 \;  \;  \land \;  \;  \Bigl\{ 0 \leq k <\frac{1}{2}-\frac{ \sqrt{1-x^2}}{(1 - 2x)}  \;  \;  \lor \;  \;  \frac{1}{2}+\frac{ \sqrt{1-x^2}}{(1 - 2x)}<k\leq 1  \Bigr\} .
\label{lam-x} 
\end{equation}
As expected  Eqs. (\ref{lam-k}) and (\ref{lam-x}) are identical to Eqs.(\ref{2kx>}) and (\ref{2xk>}). 

While this method is more direct to determine the domain of a tipping point dynamics, it does not identified the two associated  alternating attractors. The double update equation is required to localize them.

\subsection{Size 4} 

\begin{figure}
\hspace{-0.5cm}
\includegraphics[width=0.5\textwidth]{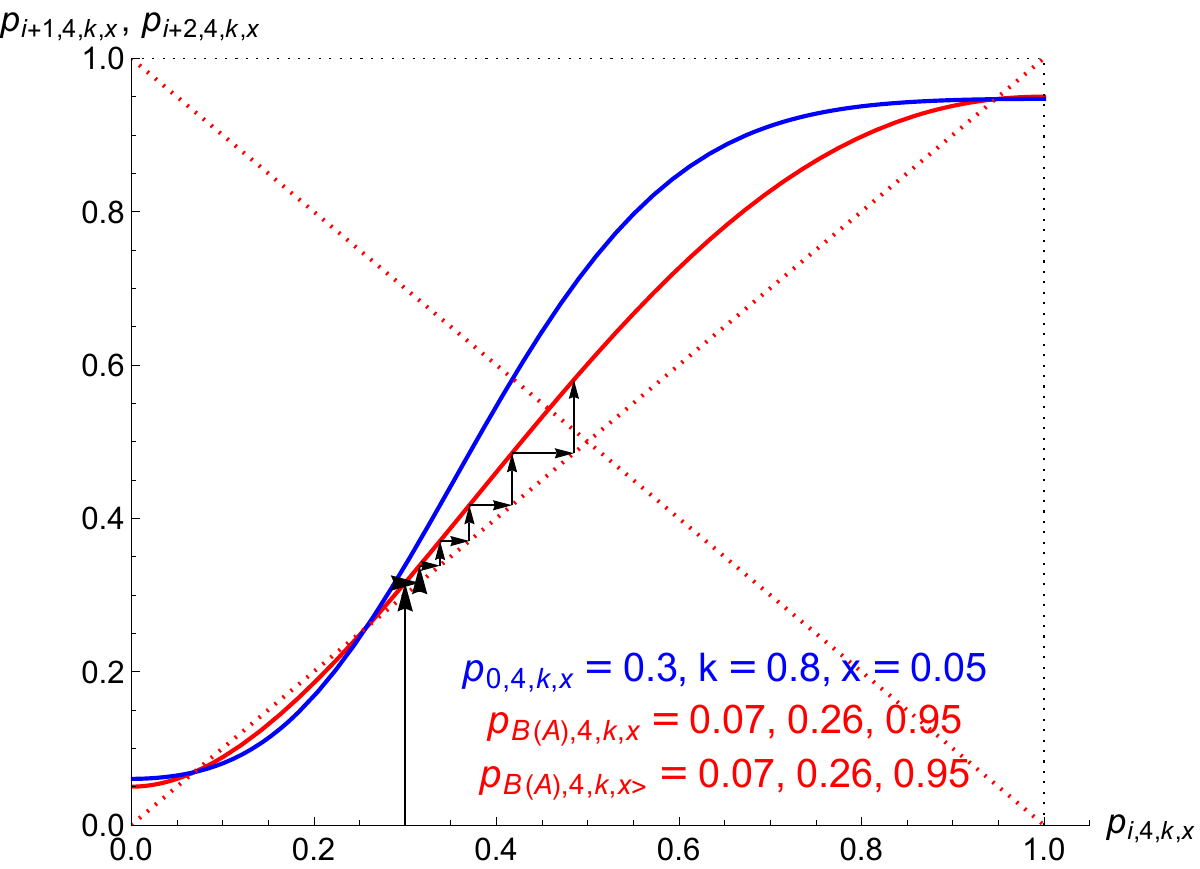} \quad
\includegraphics[width=0.5\textwidth]{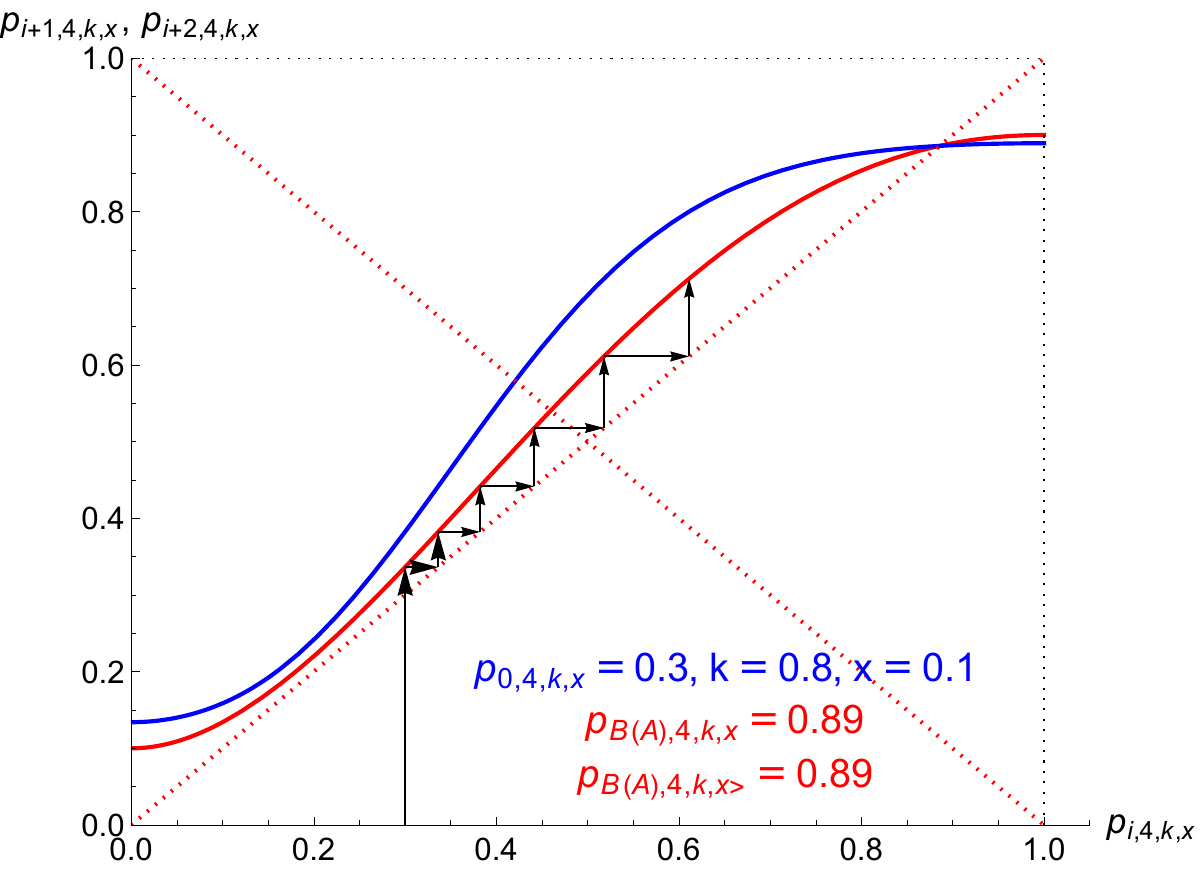}

\vspace{0.5cm}
\hspace{-0.5cm}
\includegraphics[width=0.5\textwidth]{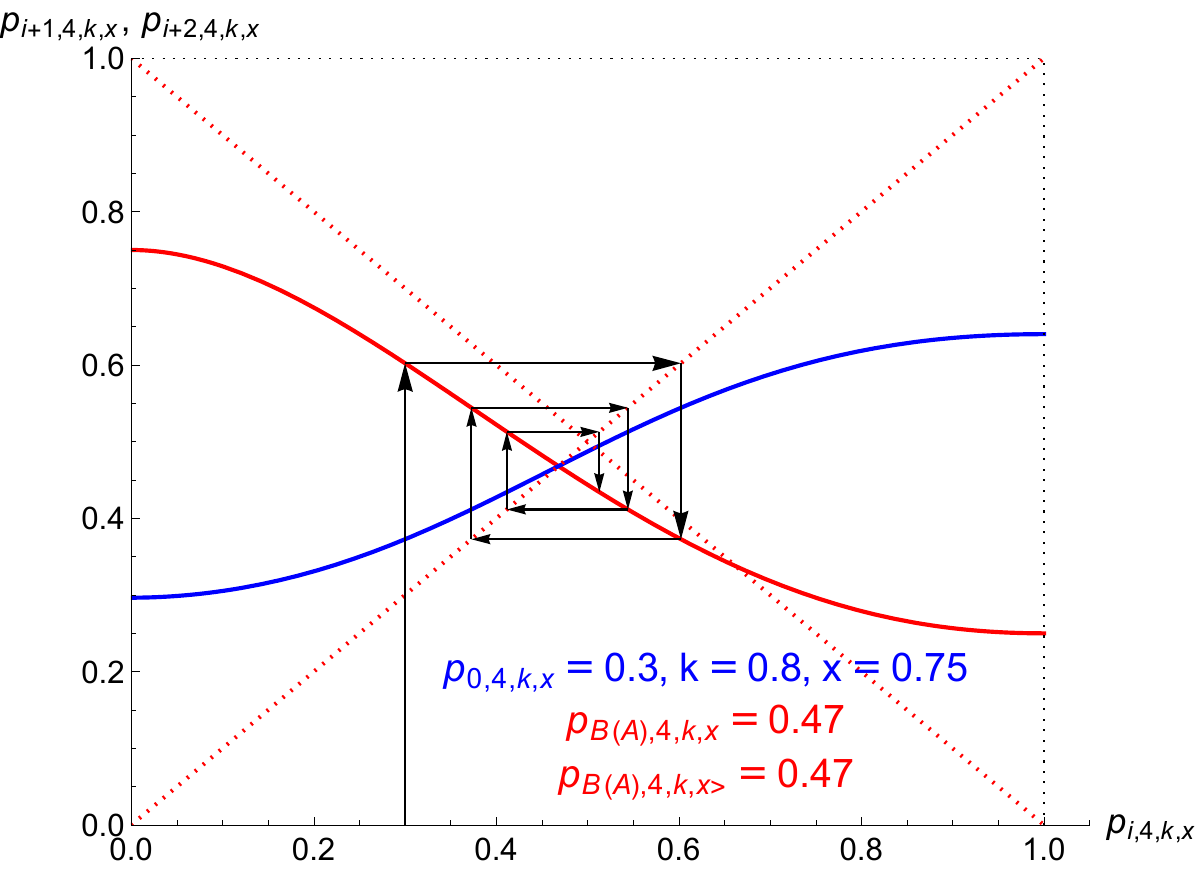}\quad
\includegraphics[width=0.5\textwidth]{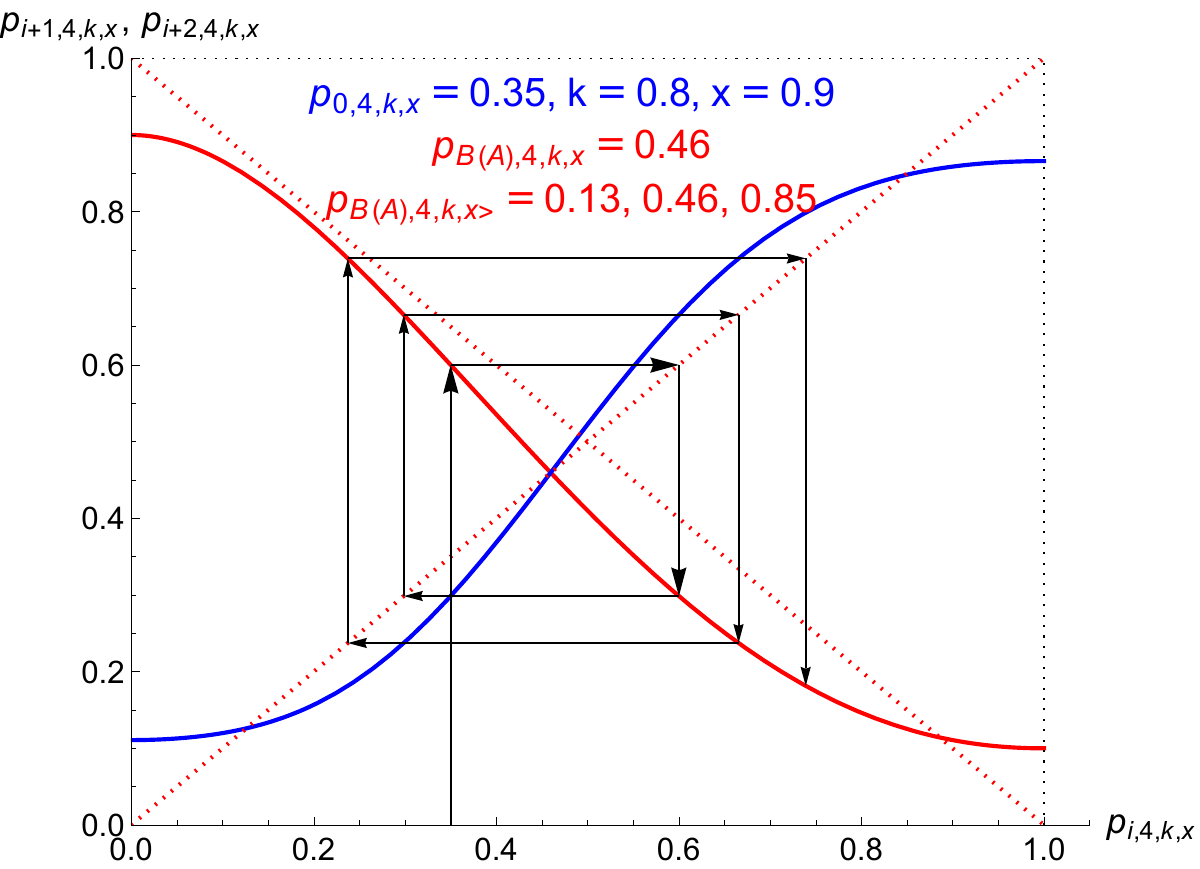}

\caption{Four different regimes of the update $p_{i+1,4,k,x}$ are shown in red as a function of both $k$ and $x$.  The blue line shows $p_{i+2,4,k,x}$, which identifies the alternating attractors. The upper left part of the Figure shows a tipping dynamics at very low proportion of contrarians with $x=0.05$ and $k=0.80$. The upper right part shows that already at $x=0.10$ the contrarians turns the dynamics to a single attractor dynamics with the attractor located at vey high value with $p_{A,4,0.8,0.1}=0.89$. At high concentrations of contrarians the dynamics stays single attractor as shown in the lower left part of the Figure with $x=0.75$ but with a lower value $p_{A,4,0.8,0.75}=0.47$.  At $x=0.85$ the dynamics becomes alternating with two attractors as exhibited in the lower right part.}
\label{u1234}
\end{figure}

Including contrarians and prejudice tie breaking turns Equation Eq. (\ref{p14x}) to,
\begin{equation}
p_{i+1,4,k,x}= (1-2x)  \left\{ p_{i,4,k,x}^4 + 4p_{i,4,k,x}^3(1-p_{i,4,k,x})+6 k p_{i,4,k,x}^2 (1 - p_{i,4,k,x})^2   \right\}  + x  ,
\label{p1kx2} 
\end{equation}  
whose fixed points expressions are not reproduced here due to very heavy analytical expressions. Moreover, above results for $x>\frac{1}{2}$ indicate that the fixed point equation $p_{i+2,4,k,x}=p_{i+1,4,k,x}$ must also be solved, i.e., a polynomial of degree 16. The associated fixed points will then be located by numerical solving.
I thus identified 4 different regimes as a function of both $k$ and $x$ as illustrated in Figure (\ref{u1234}).

For very low values of the proportion $x$ of contrarians, a tipping dynamics prevails as seen in the upper left part of the Figure. But already for $x=0.10$ a single attractor dynamics is taking place as seen in the upper right part of the Figure.  

When $x>\frac{1}{2}$ oscillations drives the dynamics with two distinct regimes. A single attractor regime takes place for  $\frac{1}{2}<x<x_c$ (lower left part of the Figure) against an alternating tipping point dynamics for   $x>x_c$ (lower right part of the Figure) where $x_c$ is very high being larger or equal to $0.85$.

\begin{figure}
\hspace{-0.5cm}
\includegraphics[width=0.5\textwidth]{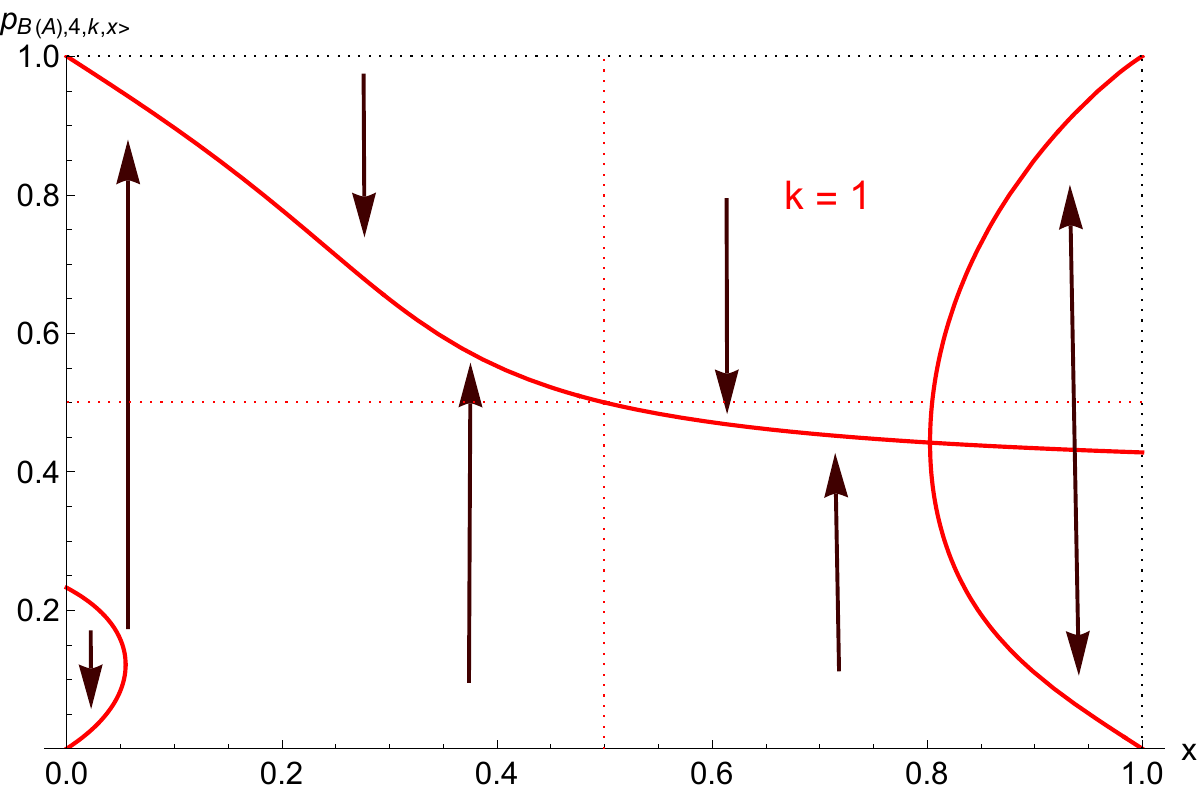} \quad
\includegraphics[width=0.5\textwidth]{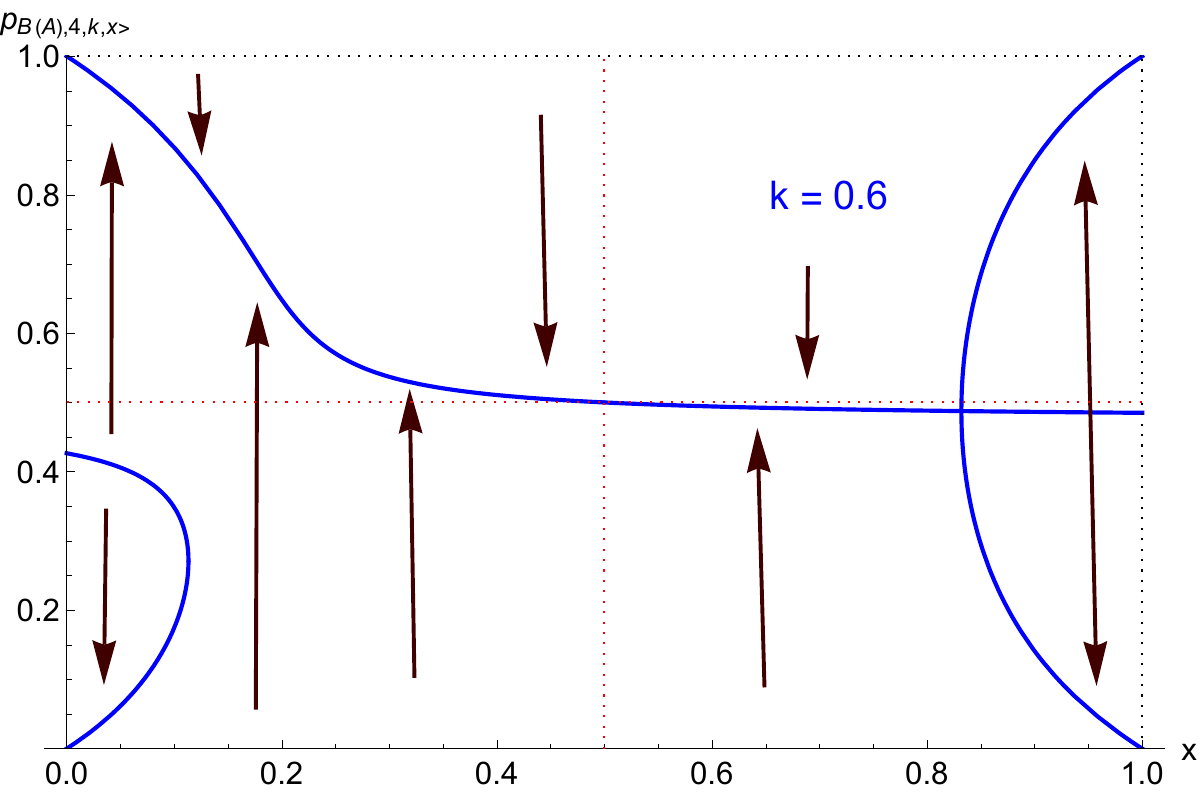}

\vspace{0.5cm}
\hspace{-0.5cm}
\includegraphics[width=0.5\textwidth]{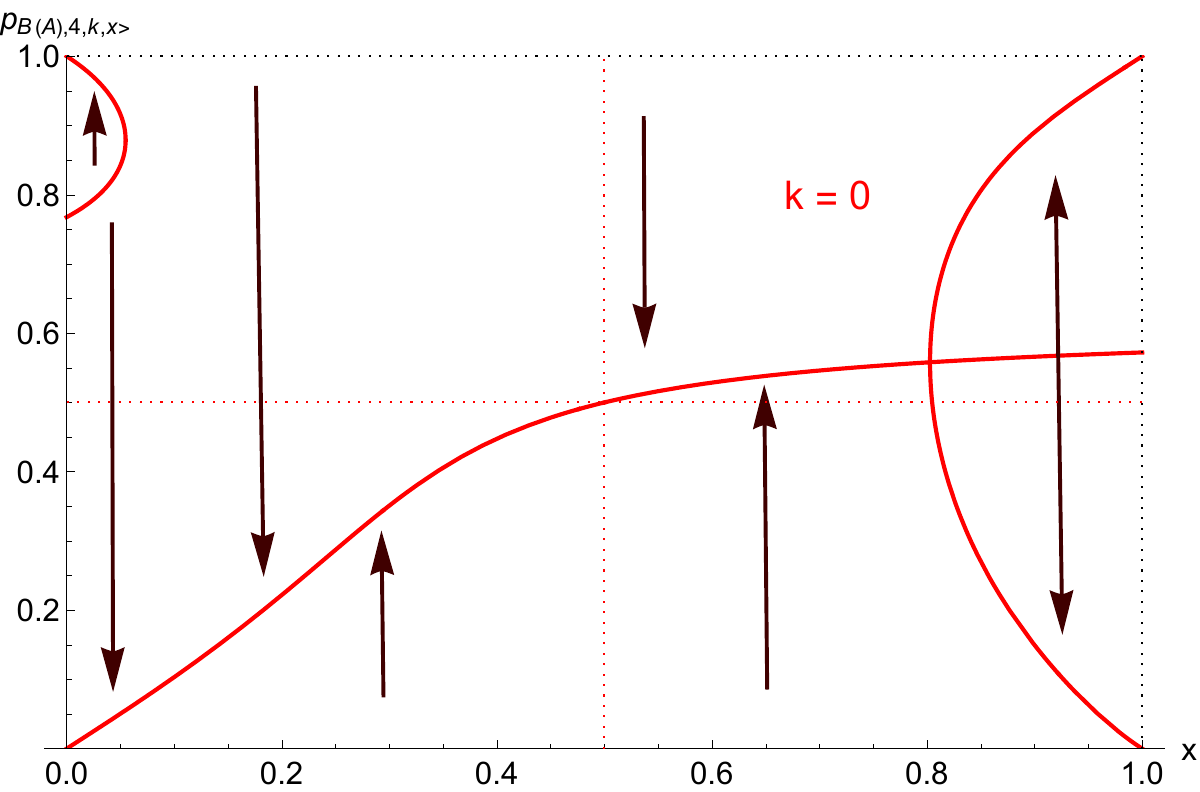}\quad
\includegraphics[width=0.5\textwidth]{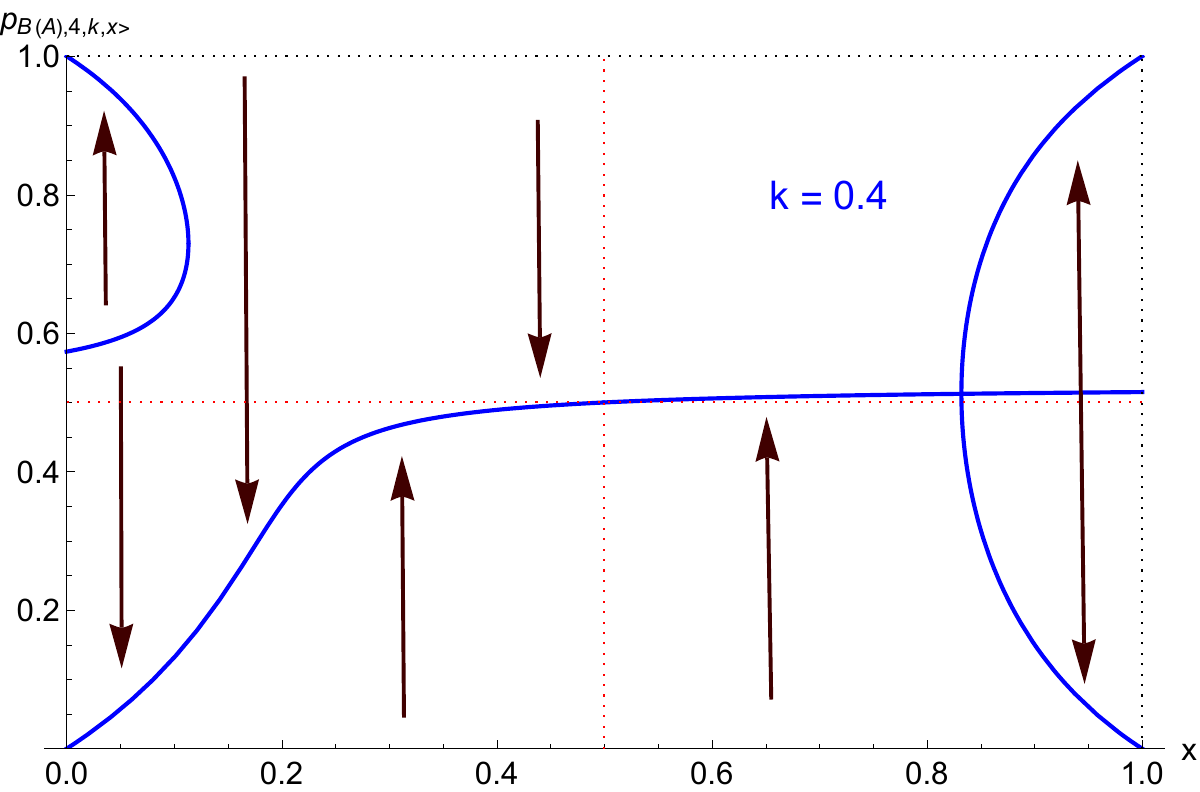}

\caption{Evolution of attractors and tipping points as a function of the proportion $x$ of contrarians for respectively $k=1$ (upper left part),$k=0$ (lower left part), $k=0.60$ (upper right part), $k=0.40$ (lower right part).  Closed curves represent tipping points and attractors while single curves denote attractors besides when inside a closed curve. Arrows indicate the direction of the flow of opinion dynamics while double arrows signal an alternating dynamics.}
\label{z1234}
\end{figure}

\begin{figure}
\hspace{-0.5cm}
\includegraphics[width=0.5\textwidth]{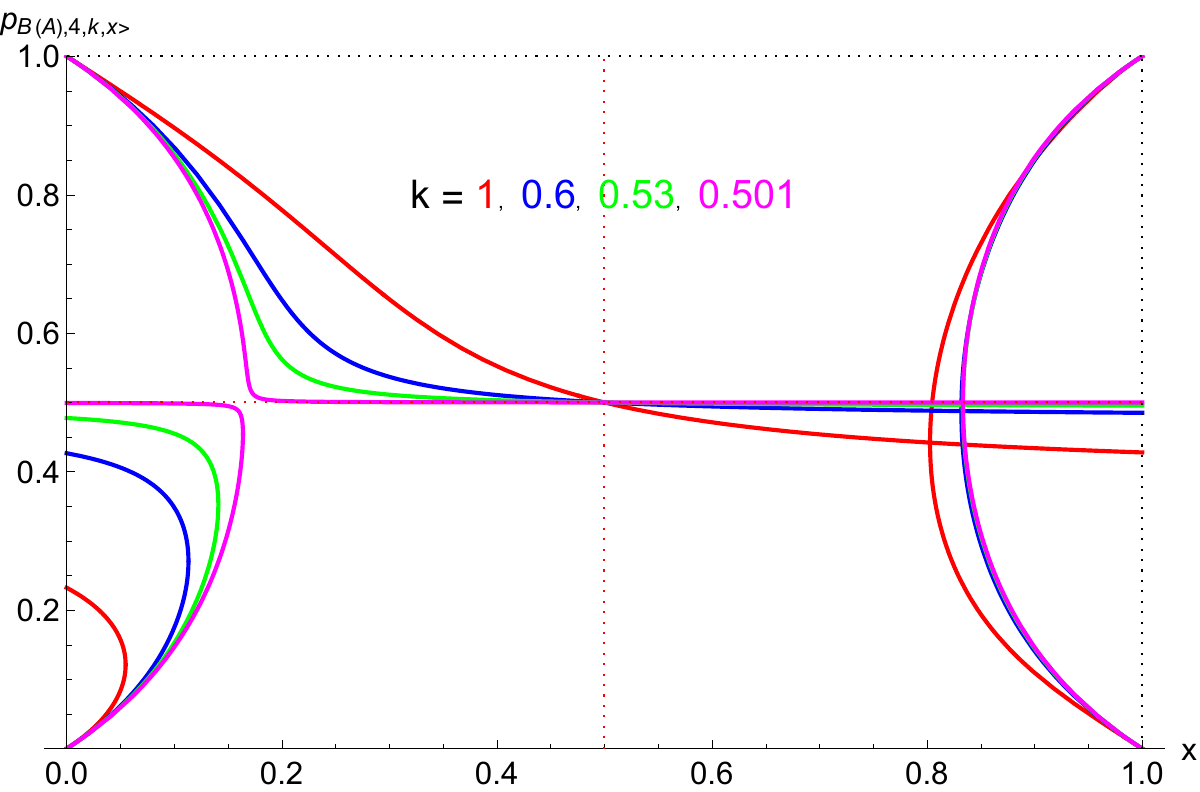} \quad
\includegraphics[width=0.5\textwidth]{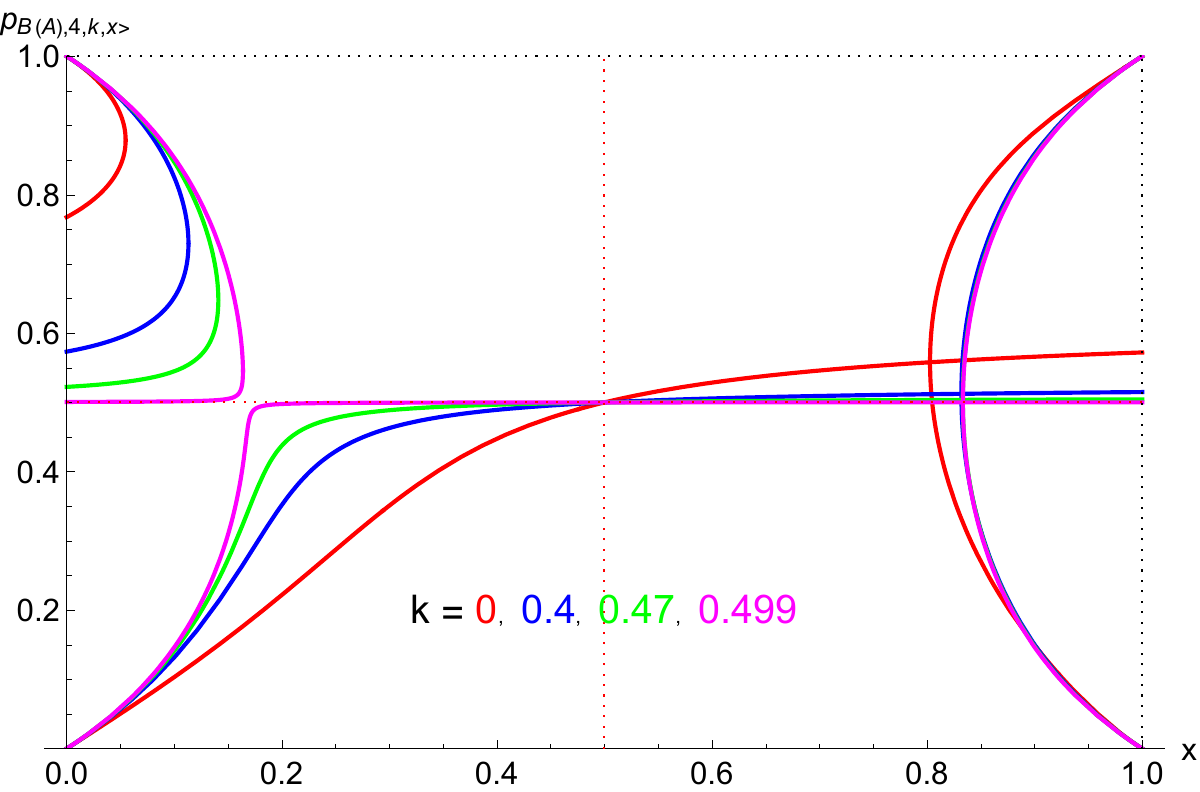}

\vspace{0.5cm}
\centering
\includegraphics[width=0.9\textwidth]{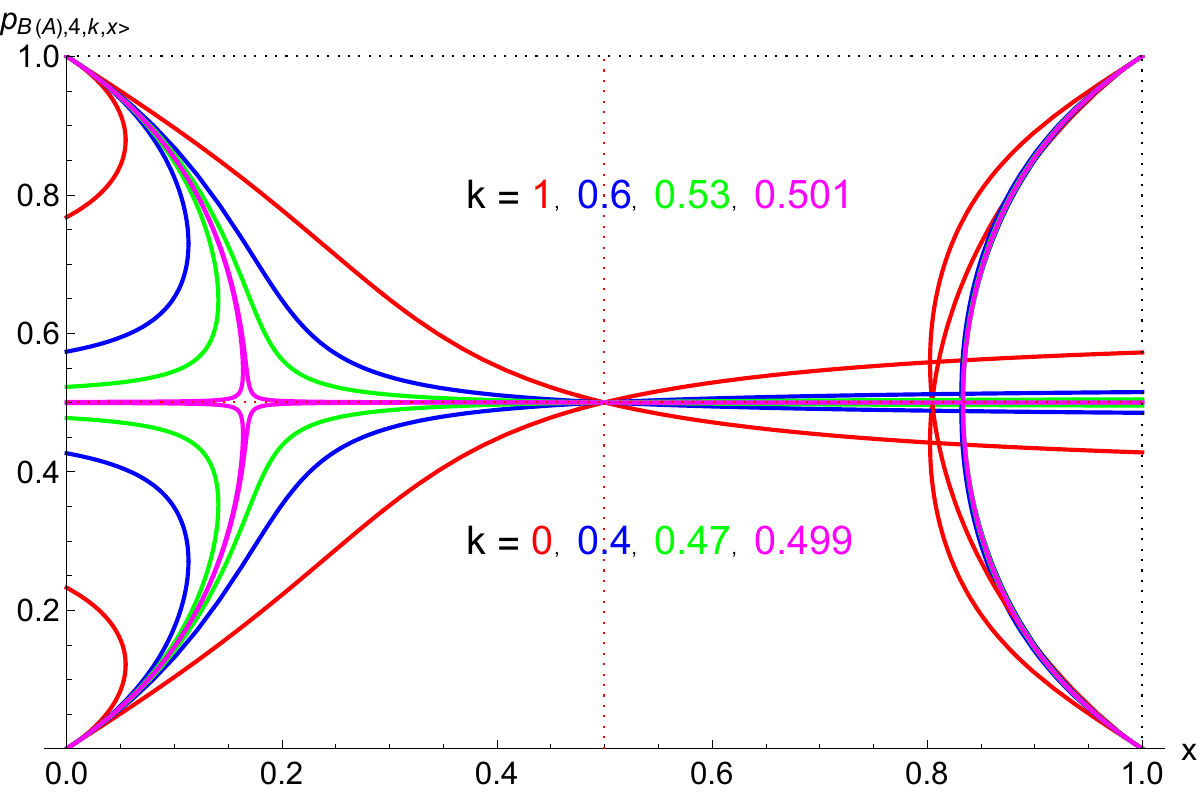} \quad
\caption{Evolution of attractors and tipping points as a function of the proportion $x$ of contrarians for respectively $k=1, 0.6, 0.501$  (upper left part) and $k=0.499, 0.4, 0$ (lower left part).  Closed curves represent tipping points and attractors while single curves denote attractors besides when inside a closed curve. Arrows indicate the direction of the flow of opinion dynamics while double arrows signal an alternating dynamics. The lower part combines both upper cases. The asymmetry between   $k<\frac{1}{2}$ and $k>\frac{1}{2}$ as well as between $x<\frac{1}{2}$ and $x<\frac{1}{2}$ is clearly seen.}
\label{z567}
\end{figure}

To build the complete landscape of the dynamics I show in Figures (\ref{z1234}) and (\ref{z567}) the evolution of attractors and tipping points as a function of $x$ for a given $k$. In the first Figure, four cases illustrate the various types of driving  landscape with respectively $k=1,0.60, 0.40, 0$. The second Figure exhibits the combined cases $k=1,0.60, 0.53, 0.501$, $k=0.499, 0.47, 0.40, 0$,  and $k=1,0.60, 0.53, 0.501, 0.499, 0.47, 0.40, 0$. The Figures shed light on the instrumental asymmetry between $k<\frac{1}{2}$ and $k>\frac{1}{2}$ as well as between $x<\frac{1}{2}$ and $x<\frac{1}{2}$.  The part $x<\frac{1}{2}$ has been previously obtained \cite{fake}.

\section{A new unexpected regime of stationary alternating polarization}

The combined results obtained for $x>\frac{1}{2}$ in the three cases $r=2, 3, 4$ validate the existence of a new unexpected regime of stationary alternating polarization, which occur only at very high proportions of contrarians. That discovery came as a surprise with such a regime overlooked up to now.

For $r=4$, iIn the range $0\leq x < x_c$ the dynamics obeys a tipping point regime with values of $x_c$ being very low as seen in Table (\ref{tt}). However, a proportion of contrarians in the range $x_c \leq  x < \frac{1}{2}$ turns the tipping point regime into a single attractor regime. 

When $\frac{1}{2}< x \leq x_{c>}$ with $x_{c>}$ being very large as seen  in Table (\ref{tt}), the dynamics stays within a single attractor regime but not now reaching the single attractor is achieved with an alternate dynamics.

Moreover, for $ x > x_{c>}$, surprisingly,  the dynamics turns back to a tipping regime but now with a stationary alternating regime between two attractors. Values of these attractors are given in Table (\ref{tt}). A symmetry between $k$ and $(1-k)$ is observed. 

\begin{figure}
\centering
\includegraphics[width=0.8\textwidth]{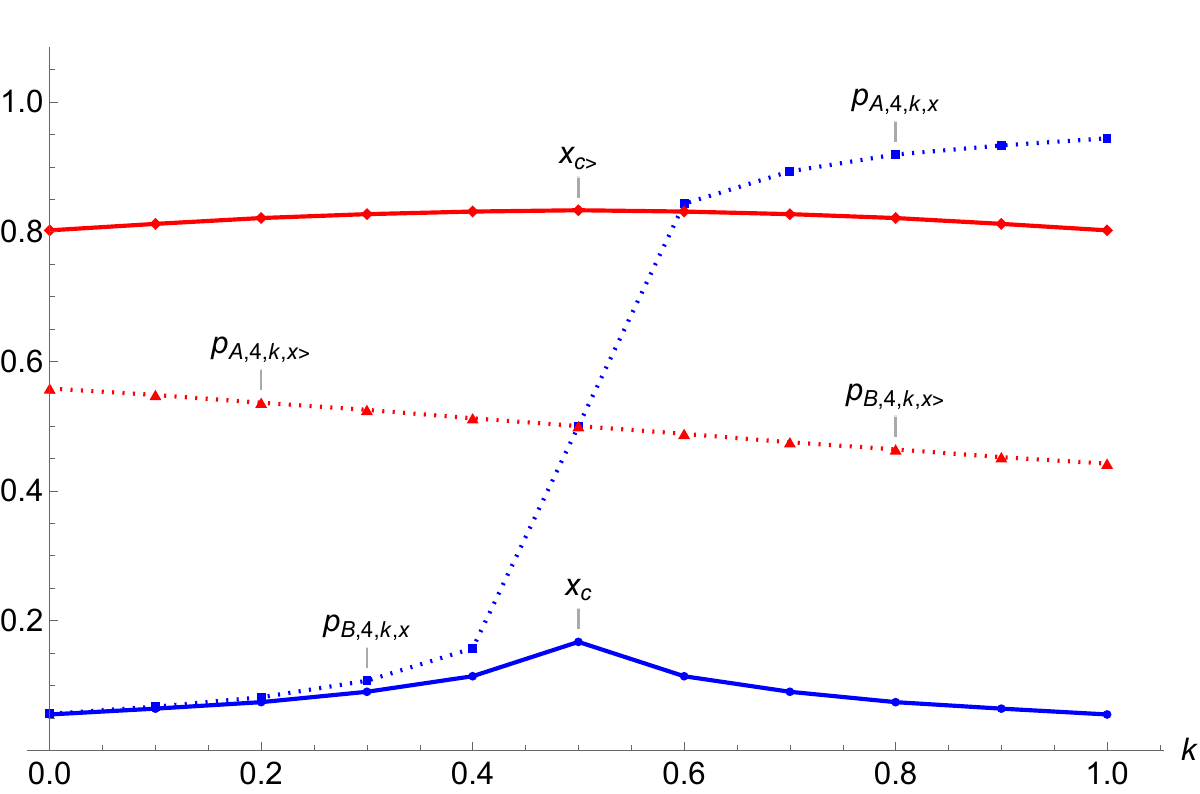} \quad
\caption{Plots of the values $x_c$, $x_{c>}$, $p_{B, k\leq 0.5, x_c}$, $p_{A, k\geq 0.5, x_c}$, $p_{A, k\leq 0.5, x_{x>}}$ and $p_{B, k\geq 0.5, x_{x>}}$  as a function of $k$ for $r=4$ according to the values of Table (\ref{tt}).}
\label{t}
\end{figure}

Figure (\ref{t}) exhibits the values $x_c$, $x_{c>}$, $p_{B, k\leq 0.5, x_c}$, $p_{A, k\geq 0.5, x_c}$, $p_{A, k\leq 0.5, x_{x>}}$ and $p_{B, k\geq 0.5, x_{x>}}$  as a function of $k$ according to the values of Table (\ref{tt}). It is worth noticing that $x_{c>} \geq 0.802$ and  $x_{c} \leq 0.167$.

Therefore, for any given values of $k$ there exists two values of $x$, respectively $x_{x}$ and $x_{x>}$, for which the regime is single attractor like. 

it is noticeable that when $k< \frac{1}{2}$ opinion B always wins with the single attractor $p_{B, k, x}$ located at low values as seen in Figure (\ref{t}). At opposite, as soon $k> \frac{1}{2}$ opinion A always wins with $p_{A, k, x}$ located at high values. A jump occurs at $k= \frac{1}{2}$. 

In contrast, opinion A always wins when $k< \frac{1}{2}$ but with a small margin with $p_{A, k, x}$ slightly above $\frac{1}{2}$. When $k> \frac{1}{2}$, B opinion wins yet with a small margin, the attractor $p_{B, k, x}$ being located slightly below $\frac{1}{2}$. Moreover, there is no jump in the attractor values, which smooth changes as a function of $k$.

\begin{table}
\begin{center}
\begin{tabular}{|c|c|c|c|c|c|c|c|c|c|c|} 
 \hline
 k & 0, 1 & 0.1, 0.9 & 0.2, 0.8 & 0.3, 0.7 & 0.4, 0.6 & 0.5 \\ 
 \hline
 \hline
$x_c$ & 0.055& 0.064& 0.074& 0.09& 0.114& 0.167 \\
 \hline
$p_{B, k\leq 0.5, x_c}$ & 0.056& 0.067& 0.0815& 0.107& 0.157& 0.500  \\
\hline
$p_{A, k\geq 0.5, x_c}$ & 0.944& 0.933& 0.919& 0.893& 0.843& 0.500  \\
\hline
\hline
$x_{c>}$ & 0.802& 0.812& 0.821& 0.827& 0.831& 0.833 \\
 \hline
$p_{A, k\leq 0.5, x_{c>}}$ & 0.558& 0.548& 0.536& 0.525& 0.512& 0.500  \\
\hline
$p_{B, k\geq 0.5, x_{c>}}$ & 0.488& 0.475& 0.464& 0.452& 0.442& 0.500 \\
\hline
\end{tabular}
\end{center}
\caption{Values of $x_c$ and $x_{c>}$ as a function of $k=0, 0.1, 0.2, 0.3, 0.4, 0.5, 0.6, 0.7, 0.8, 0.9, 1$ for $r=4$. The associated values $p_{B, k\leq 0.5, x_c}$, $p_{A, k\geq 0.5, x_c}$, $p_{A, k\leq 0.5, x_{x>}}$ and $p_{B, k\geq 0.5, x_{x>}}$  are also given.  By symmetry values are identical for $k$ and $(1-k)$.}
\label{tt}
\end{table}

\section{Conclusion}

I have built the three full three-dimensional $(p_0, k, x)$ landscapes of opinion dynamics using the Galam Majority Model (GMM) for the three sizes of local discussing groups $r=2, 3, 4$ as a function of  $0\leq p_0 \leq 1$, $0\leq k \leq 1$ and $0\leq x \leq 1$, which are the respective proportions of initial agents supporting opinion A, unavowed tie prejudice breaking in favor of opinion A and contrarians.

The GMM articulates iterative local majority rules with successive reshuffling of agents within a given social community. However, in case of a local tie the group selects by chance, i.e., without justification the opinion A with probability $k$ and opinion B with probability $(1-k)$.

For every given pair of values $(k, x)$, the dynamics in $p$ is driven by either a tipping point and two associated attractors or by a single attractor. 

In the first case, opinion A (B) needs to gather a proportion of initial support larger than the tipping point to ensure a democratic victory over time with on going discussions among the agents. However, the results have shown that this regime arises only for extreme proportions of contrarians, either very low or very high values. For $r=2$, one percent of contrarians suppresses the tipping point regime. An alternating tipping regime is obtained from 92 percent of contrarians with very low and very high values of $k$. That regime is thus very rare. For $r=4$ the tipping regime holds only for $x<0.055$ and $x>0.802$ at $k=0,1$ as seen from Table (\ref{tt}).

In the the second case, which has appeared to be the most common, the outcome of the dynamics is predetermined from the start. Opinion A (B) cannot change the outcome, either victory or defeat depending on the current location of the single attractor with respect to $50\%$.

The results have thus indicated that the expected democratic character of free opinion dynamics is rarely satisfied. Indeed, most part of the subspace $(k, x)$ of the three dimensional space $(p_0,k,x)$ is governed by a single attractor regime. Therefore, any initial supports of opinions A and B end up to the unique attractor, which is either above or below fifty percent for opinion A (B). The final outcome of a democratic public debate is thus predetermined independently of which opinion started being majority in the related community.

On this basis, the only potential option for the supporters of the predetermined defeated  opinion, would be trying to modify the actual pair $(k, x)$ to reach a location, which is beneficial to that  opinion. Nevertheless, the challenge to identify and set the means to implement changes in $k$ or and $x$ is out of the scope of the present paper.

\subsection*{A word of caution}

Last but not least,  a word of caution is in order here: in case, groundbreaking methods to actually change $k$ or and $x$ are pionnered,  I am aware of the of risk  that my work could become an effective tool to manipulate opinion dynamics. However, the current situation, provided my model is sound, is shown to be embedded with  invisible mechanisms, which are likely to twist``naturally" the democratic balance of a public debate in most cases. That highlights a kind of unconscious self-manipulation which skews the initial majority's will. People can thus be taken to opposite wishes leading to possible social and political disasters.

I am convince that only the discovering of the laws governing human behavior, in particular the dynamics of public opinion, could extract people as a community, from being driven by their current ignorance and wrong beliefs to make inappropriate choices.

In addition, once validated, a hard science implies both predicting and acting upon related phenomena. We are on a promising track with still a long way to go.

At this stage, my ethical responsibility as a scientist is to ensure that my results are freely accessible to everyone. The responsibility of the consequences of their use then lies among the future users, policy makers and others.

\end{document}